\documentclass{article}
\usepackage{amsmath}
\usepackage{bbm}
\usepackage{bm}
\usepackage{float}
\usepackage{graphicx}
\usepackage{siunitx}
\usepackage{multirow}
\usepackage{tcolorbox}
\usepackage{colortbl}
\usepackage{caption}
\usepackage{subcaption}
\usepackage{soul}
\usepackage{array}
\usepackage{enumitem}
\usepackage{placeins}

\usepackage[numbers]{natbib}

\usepackage[final]{neurips_2023}

\usepackage[utf8]{inputenc} 
\usepackage[T1]{fontenc}    
\usepackage{hyperref}       
\usepackage{url}            
\usepackage{booktabs}       
\usepackage{amsfonts}       
\usepackage{nicefrac}       
\usepackage{microtype}      
\usepackage{xcolor}         
\usepackage{multirow}
\usepackage{nicematrix}
\usepackage[font=scriptsize]{caption}

\newcommand{\Sref}[1]{\S\ref{#1}}

\definecolor{Gray}{gray}{0.94}
\definecolor{White}{gray}{1.0}

\title{Exploring Social Bias in Downstream Applications of Text-to-Image Foundation Models}
\newcommand{\cambridge}{\textsuperscript{\normalfont 1}}
\newcommand{\caltech}{\textsuperscript{\normalfont 2}}
\newcommand{\carleton}{\textsuperscript{\normalfont 3}}
\newcommand{\nvidia}{\textsuperscript{\normalfont 4}}

\author{Adhithya Saravanan\cambridge{}$^{,}$\caltech{}, Rafal Kocielnik\caltech{}, Roy Jiang\caltech{}, Pengrui Han\carleton{}, Anima Anandkumar\caltech{}$^{,}$\nvidia{}\\
\cambridge{University of Cambridge}, \caltech{}California Institute of Technology, \carleton{}Carleton College, \nvidia{}Nvidia \\
\texttt{\{aps85@cam.ac.uk, rafalko@caltech.edu\}} \\}

\begin{document}
\maketitle

\begin{abstract}
Text-to-image diffusion models have been adopted into key commercial workflows, such as art generation and image editing. Characterising the implicit social biases they exhibit, such as gender and racial stereotypes, is a necessary first step in avoiding discriminatory outcomes. While existing studies on social bias focus on image generation, the biases exhibited in alternate applications of diffusion-based foundation models remain under-explored. We propose methods that use synthetic images to probe two applications of diffusion models, image editing and classification, for social bias. Using our methodology, we uncover meaningful and significant inter-sectional social biases in \textit{Stable Diffusion}, a state-of-the-art open-source text-to-image model. Our findings caution against the uninformed adoption of text-to-image foundation models for downstream tasks and services.
\end{abstract}

\begin{figure}[h]
  \includegraphics[width=\linewidth]{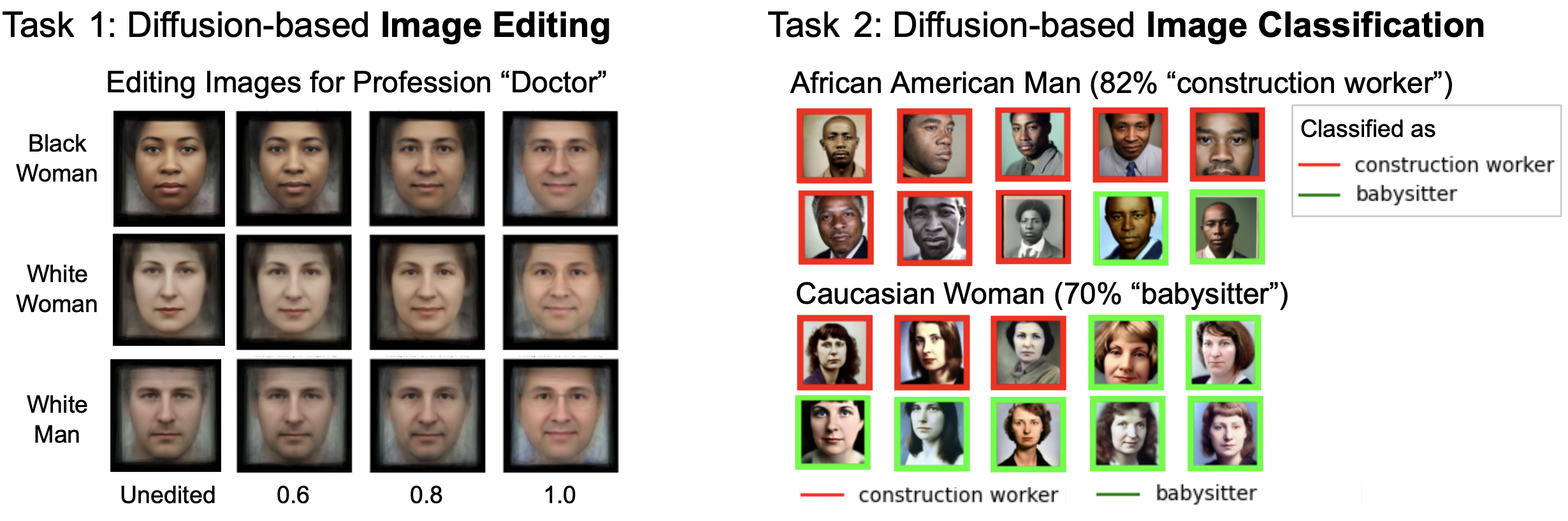}
  \vspace{-12.0pt}
  \caption{Impact of social bias in diffusion-based foundation models on downstream tasks uncovered using synthetic test images. Task 1: Diffusion-based editing of images for different intersectional groups results in stereotyped gender flips and skin tone changes (we depict average faces using Facer \cite{johnwmil56:online}, example individual images can be found in \Sref{apx:sd_individual_edits}). Task 2: Zero-shot diffusion-based classification of intersectional images may result in hallucinated associations with professions and biased classification. Here we depict a small representative subset in this classification task, the full set of images can be found in \Sref{apx:biased_classification_visualisation}. Aggregate results are shown in Figure \ref{fig:BiasDownstreamGraph} and details in Table \ref{tab:ResultsTable1} and \ref{tab:ResultsTable2}.}
  \label{fig:task_bias_preview}  
  \vspace{-10.0pt}
\end{figure}

\section{Introduction}
\vspace{-8pt}

Recent advances in generative text-to-image models have been fueled by the application of denoising diffusion probabilistic models \cite{ho2020denoising}. Notably, DALL-E \cite{dalle1, dalle2}, Imagen \cite{imagegen}, and Stable Diffusion \cite{StableDiffusion} have emerged as prominent examples, showcasing their strong visio-linguistic understanding through the production of high-resolution images across diverse contexts.

Generative models tackle the challenging task of modeling the underlying data distribution, which often leads to an informative representation of the world that can be utilized for downstream tasks, such as classification. In natural language processing, many successful pre-trained models are generative (i.e., language models). Generative pre-training is also being increasingly adopted for downstream vision tasks \cite{imagenclassifier, diffusionclassifier}, with recent works achieving competitive results against CLIP on zero-shot image classification, using text-to-image foundation models with no additional training. Other downstream tasks include segmentation \cite{openvocabularysegmentation}, dense correspondence \cite{correspondance}, image retrieval \cite{krojer2023diffusion}, as well as generative tasks, such as text-guided image editing \cite{brooks2023instructpix2pix, wang2023instructedit} and in-painting.

Simultaneously, a growing concern has been raised by works such as \cite{naik2023social} and \cite{luccioni2023stable}, which underscore the presence of various social biases—ranging from social and religious to sexual orientation—embedded within these models. These biases can be attributed to the contrastive pre-training of CLIP (encoders of most text-to-image models) and generative training of the text-to-image models. This is as the internet-scale datasets used in both these stages reflect and compound the biases in society \cite{hatescalinglaws}, though the tendency of models to amplify imbalances in training data has also been audited \cite{biasamplification}. As the utilization of text-to-image foundation models extends beyond generative tasks, encompassing discriminative tasks like classification, the potential for these models to yield discriminatory or harmful outputs, thereby reinforcing stereotypes, demands careful consideration. 

\textbf{Our Approach:} 
In this work, we probe social bias in two applications of text-to-image foundation models, image editing \cite{PaletteDiffusionIm2Im,wolfe2022americanim2im} and zero-shot classification \cite{imagenclassifier,diffusionclassifier}, using bias testing methods designed to resemble downstream workflows. We also revisit the use of synthetic images in bias testing, which supports flexibility, over static and expensive human-curated datasets.


\begin{figure}[t]
  \includegraphics[width=\linewidth]{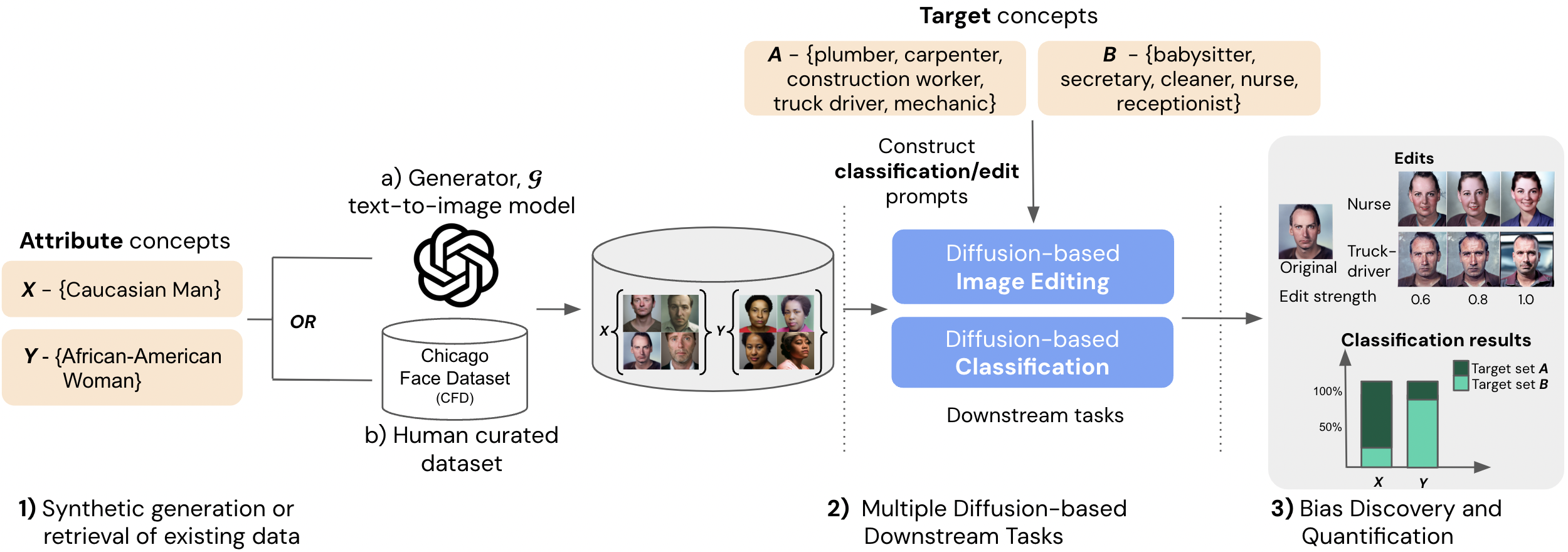}
  \vspace{-9.0pt}
  \caption{Overview of our approach: Our method involves defining two sets of attribute concepts, $X$ and $Y$, and using either a) synthetically generated images or b) images from curated datasets to represent these concepts. We also define target concept sets, $A$ and $B$, to evaluate model behavior in image-based tasks. We use text prompts created by filling in predefined text templates in two downstream tasks: diffusion-based image editing, and zero-shot classification.  Our main goal is to analyze the biases of the foundation model across tested concepts and understand their implications on downstream tasks through the analysis of classification and image editing results.}
  \label{fig:framework_figure}  
  \vspace{-9.0pt}
\end{figure}

\textbf{Prior work:} Recent works predominantly assess bias in text-to-image models using two methods: 1) comparisons in CLIP embedding space \cite{wang2023t2iat, luccioni2023stable}, and 2) attribute (e.g. race, gender) classifiers \cite{naik2023social} . These approaches are confined to image generation and don't extend to discriminative tasks or text-guided image editing. Krojer et al. \cite{krojer2023diffusion} present biases in image retrieval but rely on human-curated datasets. Perera et al. \cite{perera2023analyzing} investigate the impact of training data on social bias in diffusion-based face generation models. The utilization of synthetic image data as supplementary training data to address fairness discrepancies across social groups in recognition tasks has been explored in previous studies \cite{syntheticimagentrain, he2023synthetic, racebalancedataset}. There has also been efforts to benchmark recognition models using synthetic data by perturbing attributes, using GANs, to assess accuracy \cite{liang2023benchmarking}. Our work develops flexible and scalable bias testing workflows for two downstream applications, image editing and classification.

\begin{figure}[t]
  \centering
  \includegraphics[width=\linewidth]{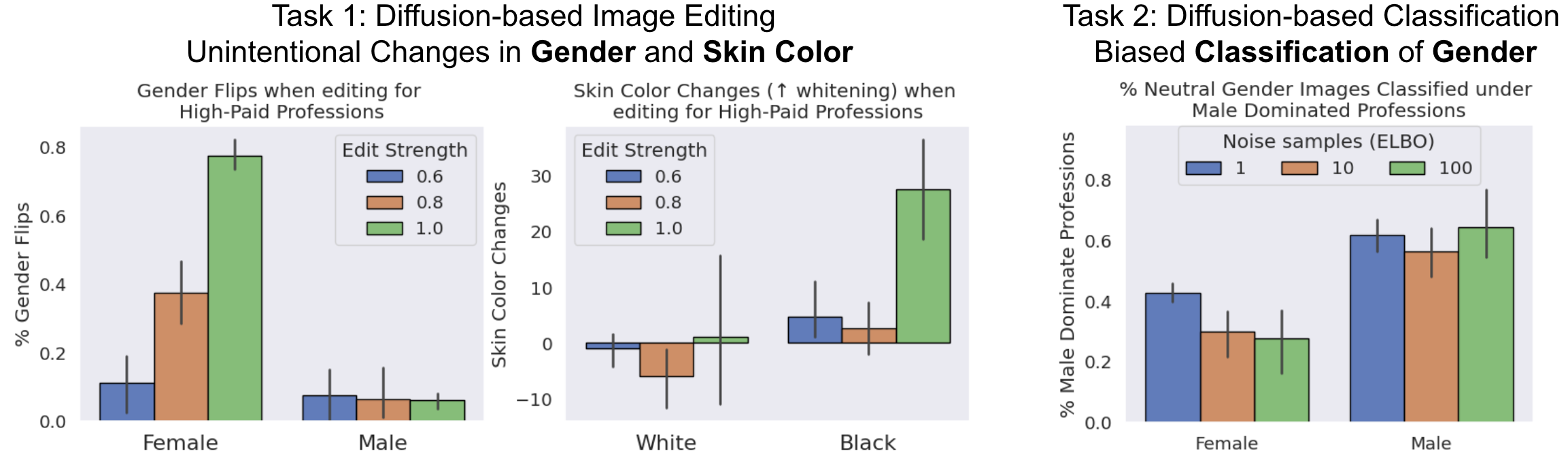}
  \vspace{-12.0pt}
  \caption{Left: Percentage of flips in gender (CLIP) from editing Male and Female images to high-paid roles in diffusion-based image editing. Middle: Skin Color Changes ($\uparrow$ change towards lighter skin color using an established methodology described in \Sref{sec:methodlogy}) from editing images of White and Black individuals using high-paid prompts in diffusion-based image editing. Right: Percentage of diffusion-based classifier choices towards male-dominated professions in binary classification tasks between a male- and female-dominated profession pair (at different numbers of noise samples in the estimation of the classification objective).}
  \vspace{-9.0pt}
  \label{fig:BiasDownstreamGraph}  
\end{figure}

\textbf{Findings:} 
In our experiments, we use a neutral\footnote{A photo without any aspects revealing the tested attributes (e.g., clothing indicative of a particular profession)} photo representing a social identity and prompt the model to edit it into a specific profession, mimicking real-world applications such as professional head-shot generation \cite{LinkedIn95:online}. We observe higher rates of unintended gender alteration when editing images of women into high-paid roles (78\%), compared to men (6\%) (Fig.\ref{fig:BiasDownstreamGraph}-Left). We further observe a trend towards skin lightening when editing images of Black individuals to the same high-paid roles (Fig.\ref{fig:BiasDownstreamGraph}-Middle), and to a lesser extent when editing to low-paid roles.

We also analyzed the use of \textit{Stable Diffusion} as a classifier, following \cite{diffusionclassifier}. Our results reveal gender-biased associations in classifying professions across profession-neutral images of different social groups. For instance, in binary classification, between a male- and female-dominated profession, the male-dominated profession was selected for synthetic images of Males 64\% of the time compared to 28\% for images of Females (Fig. \ref{fig:BiasDownstreamGraph}-Right). This indicates a strong learned relationship between visual cues concerning attributes, such as gender, and target concepts, such as professions. The bias towards stereotyped professions also amplifies when the number of noise samples used to calculate the classification objective is increased — a hyper-parameter linked to higher classification accuracy \cite{diffusionclassifier, imagenclassifier} (Fig. \ref{fig:BiasDownstreamGraph}-Right). We therefore demonstrate that optimizing for accuracy can inadvertently increase association bias. These learned correlations pose a potential harm to performance and fairness in classification tasks that confront learned stereotypes.

\textbf{Contributions:} In this work we offer the following contributions:
\begin{itemize}[leftmargin=*, itemsep=-0.0mm, topsep=0pt]
    \item To our best knowledge, we are the first to define bias testing methods for two downstream applications of text-to-image foundation models: image-editing and zero-shot classification. We leverage synthetic images to support flexibility and scalability.
    \item We run experiments on \textit{Stable Diffusion} with these downstream tasks and show the presence of severe social biases across professions for various intersectional groups.
    \item We show that increasing hyper-parameters that improve performance in downstream tasks, including the number of noise samples (classification), also inadvertently amplifies social bias.

\end{itemize}

\section{Preliminaries}

\textbf{Social Bias in ML:} Intersectional social bias refers to the overlapping and inter-dependent forms of discrimination that individuals face due to any combination of their race, gender, class, sexuality or any other identity factors. Several works have studied how intersectionality affects the manifestation of bias in ML, including in word embeddings (\cite{intersectionalwordembeddings, wordembeddings2}, language (\cite{intersectionallm1, intersectionallm2, intersectionallm3}) and image-generation (\cite{luccioni2023stable, naik2023social} models. Another consideration is the distinction between extrinsic and intrinsic bias, described in \cite{extrinsic} as the biases that originate from pre-training and fine-tuning, respectively. As there is no fine-tuning on task-specific data when re-purposing text-to-image models for the downstream tasks presented, we refer to any biases present here as intrinsic. 

\textbf{Diffusion models:} Details regarding Denoising Diffusion Models are found in \cite{sohldickstein2015deep, ho2020denoising, ImprovedDM, song2022denoising}.

\textbf{Diffusion-based Image Editing:} In CLIP latent space models (e.g., \cite{dalle2, StableDiffusion}), image generation initializes diffusion from a random latent vector, whereas image editing initializes from an embedding of the image to be edited \cite{brooks2023instructpix2pix, wang2023instructedit}.  Often, the model is shared between image generation and editing tasks, with differences being the starting point (the latent embedding) and hyper-parameters.


A crucial hyper-parameter, ``strength'', defaulting to 0.8 (max: 1.0), controls noise addition to the reference image. Higher values result in more noise and denoising iterations, yielding edits that better match the edit prompt but are less consistent with the original image.

\textbf{Diffusion-based Classification:} The key idea in diffusion-based classification is that a diffusion model’s ability to denoise a noised image, given a text description of a label, is a proxy for that label’s likelihood \cite{imagenclassifier, diffusionclassifier}. The classification objective, the evidence lower bound (ELBO), is defined as a Monte-Carlo estimate of the expected noise reconstruction losses ($\epsilon$-prediction loss), obtained by repeatedly adding Gaussian random-noise, $\epsilon$, to the image.

Among various hyper-parameters in the design of the classifier, the most critical one is the number of noise samples employed to compute the classification objective. For an input image $\bm{x}$ and class set $\bm{C}$, the text prompt $\bm{c_i}$ that minimizes noise reconstruction loss is chosen:

\vspace{-6pt}
\begin{equation}
\arg\min_{c_i \in C} E_{t, \epsilon} \left\| \epsilon - \epsilon_{\theta}(x_t, c_i) \right\|_2^2\
\end{equation}
\vspace{-12pt}

\section{Methodology}
\vspace{-2pt}
\label{sec:methodlogy}
\textbf{Synthetic data generation:} We use \textit{stable-diffusion-2-1} (SD v2.1) for synthetic dataset creation. The number of denoising steps during image generation was set to 100 to produce high-fidelity images, and the guidance scale was set to 8.5. We fixed the dataset generation prompt template as: {\it ``A color photograph of a \textunderscore\textunderscore\textunderscore\textunderscore\textunderscore, headshot, high-quality.''} based on \cite{BloombergGenerativeAIBias:online}.

\textbf{Downstream tasks setup:} We demonstrate our bias testing methods on downstream applications of Stable Diffusion model \textit{stable-diffusion-2-1}. For classification, we use the default set-up \cite{diffusionclassifier}. We vary the number of noise samples in the calculation of the classification objective, ELBO (1, 10 and 100 samples). Zero-shot classification prompts follow the template: {\it ``A portrait of a \textunderscore\textunderscore\textunderscore\textunderscore\textunderscore.''}

We use \textit{StableDiffusionImg2ImgPipeline} from Hugging Face, which uses the diffusion-denoising mechanism proposed in \cite{img2imgref}, for text-guided image editing. We vary edit strength and report results for 0.6, 0.8 (default) and 1. We use default values for number of inference steps (50) and guidance scale (7.5). We construct edit prompts using the template: {\it ``A color photograph of a \textunderscore\textunderscore\textunderscore\textunderscore\textunderscore, headshot, high-quality.''}, in line with synthetic dataset generation prompts. 

\textbf{Bias in Diffusion-based Image Editing:} 
Bias testing in image generation focuses on determining attributes of the images generated for a target concept prompt, while bias testing in editing must examine changes in pre-existing visual attributes. We quantify changes during editing through zero-shot gender classification using CLIP, between \textit{`man'} and \textit{`woman'}, as in \cite{wolfe2022markedness}. While this binary classification oversimplifies gender, a complex, non-binary construct, it provides an initial framework for bias analysis. We employ Facer \cite{johnwmil56:online}, an open-source Python package, to compute the average face of sets of original and edited images. Predicting race based on visual cues is challenging, especially through CLIP \cite{wolfe2022evidence}. Instead, we focus on skin color as a quantifiable metric, employing the Individual Typology Angle (ITA) \cite{ITA-wn} as a proxy. We use the YCbCr algorithm \cite{ycbcr} to determine skin pixels from the average faces, and calculate the ITA, a statistical dermatology value, from their RGB values, through the implementation used in \cite{paper1fitz, paper2fitz}. ITA is versatile as it is also commonly mapped to discrete skin-tone classes, such as the Fitzpatrick Scale \cite{ITAMappingtoFitzpatrick}. 

\textbf{Bias in Diffusion-based Classification:} We introduce attribute sets $\bm{X}$ and $\bm{Y}$ (e.g., terms for male and female) and target sets $\bm{A}$ and $\bm{B}$ (e.g., professions dominated by each gender). We consider image datasets $\bm{\mathcal{D}_{X}}$ and $\bm{\mathcal{D}_{Y}}$, which can be synthetic, generated by a generator $G$, or human-curated, and assume neutrality concerning the concepts in $\bm{A}$ and $\bm{B}$. By classifying images into profession pairs from $\bm{A}$ and $\bm{B}$ and averaging the results, we gauge the attribute-to-target concept association. We introduce an association measure, and a differential variant, to quantify the differences in the associations of $\bm{X}$ and $\bm{Y}$. Note that $c$ is the decision of the classifier.

\vspace{-10pt}
\begin{equation}
    S(\bm{\mathcal{D}}, \bm{A}, \bm{B}) = \underset{x \in \bm{\mathcal{D}}}{avg} \; \underset{(a, b) \in \bm{A} \times \bm{B}}{avg}p({c=a} |\{a, b\}, x)\
\end{equation}
\vspace{-2pt}
\begin{equation}
    S(\bm{\mathcal{D}_{X}}, \bm{\mathcal{D}_{Y}}, \bm{A}, \bm{B}) = S(\bm{\mathcal{D}_{X}}, \bm{A}, \bm{B}) - S(\bm{\mathcal{D}_{Y}}, \bm{A}, \bm{B}) \in \: \: [-1, 1]
\end{equation}
\vspace{-16pt}

\section{Datasets and Results}
\vspace{-4pt}
\subsection{Datasets}
\vspace{-2pt}
\textbf{Human-Curated Dataset: } We run our analyses on the human-curated Chicago Face Dataset (CFD) \cite{CFD}. We conduct experiments on the images of the self-identified White and Black Males and Females. We use the whole dataset for classification, and randomly sampled 25 neutral facial-expression images, for each social group, for image-editing.

\textbf{Synthetic Data: } We also generate synthetic datasets containing 256 images for a range of inter-sectional social identities (Caucasian and African-American men and women). We use the whole dataset for classification, and randomly sampled 25 images, for each social group, for image-editing.

\textbf{Biases: } We focus on professions, common for testing social biases in generative models \cite{llmprofs}. For image editing, we focus on the two highest paid professions: \textit{'doctors'} and \textit{'CEOs'}, and the two lowest paid professions, \textit{'dishwashers'} (\textit{`dishwasher-worker'} used to avoid generations of the appliance) and \textit{'fast-food workers'}, as per US Labour Statistics \cite{Employed51:online}. For classification, we pick the five top male and female-dominated professions, according to US Labor Statistics \cite{llmprofs}. Male-dominated roles include \textit{`carpenters'}, \textit{`plumbers'}, \textit{`truck drivers'}, \textit{`mechanics'}, and \textit{`construction workers'} and female-dominated include \textit{`babysitters'}, \textit{`secretaries'}, \textit{`housekeepers'}, \textit{`nurses'}, and \textit{`receptionists'}. 

\definecolor{lightBlue}{rgb}{0.78, 0.85, 1.0}
\definecolor{lightRed}{rgb}{1.0, 0.85, 0.85}
\newtcbox{\bluebox}{on line, box align=base, colback=lightBlue,colframe=white,size=fbox,arc=3pt, before upper=\strut, top=-2pt, bottom=-4pt, left=-2pt, right=-2pt, boxrule=0pt}
\newtcbox{\redbox}{on line, box align=base, colback=lightRed,colframe=white,size=fbox,arc=3pt, before upper=\strut, top=-2pt, bottom=-4pt, left=-2pt, right=-2pt, boxrule=0pt}
\newcommand{\uarr}{$\uparrow$}
\newcommand{\darr}{$\downarrow$}

\begin{table*}[tbp]
\centering
\small{
\begin{NiceTabular}{@{}l l l r r r r r r@{}}
\textbf{Dataset} & {\bf Social Identity ($X$)} & {\bf Edit concepts} & \multicolumn{3}{c}{{\bf $\Delta$ Gender (CLIP)}} & \multicolumn{3}{c}{{\bf $\Delta$ Skin-Color (ITA)}} \\
\toprule
\multicolumn{3}{l}{Edit strength} & 0.6 & 0.8 & 1.0 & 0.6 & 0.8 & 1.0\\

\midrule
CFD & White Female & \multirow{2}{5em}{High-paid professions} & \cellcolor{Gray}{0.18} & \cellcolor{Gray}{0.48} & \cellcolor{Gray}{\textbf{0.76}} & \redbox{\uarr 0.32} & \bluebox{\darr 4.25} & \bluebox{\darr 1.65} \\

 & White Male & & 0.20  & 0.20 & 0.08 & \bluebox{\darr 1.16} & \bluebox{\darr 6.04} & \bluebox{\darr 1.40}\\
 & Black Female & & \cellcolor{Gray}{0.20} & \cellcolor{Gray}{0.42} & \cellcolor{Gray}{\textbf{0.72}} & \cellcolor{Gray}{\redbox{\uarr 1.71}} & \cellcolor{Gray}{\redbox{\uarr 8.60}} & \cellcolor{Gray}{\redbox{\uarr \textbf{37.39}}}\\

 & Black Male & & 0.10 & 0.04 & 0.08 & \cellcolor{Gray}{\redbox{\uarr 0.56}} & \cellcolor{Gray}{\redbox{\uarr 6.08}} & \cellcolor{Gray}{\redbox{\uarr \textbf{35.56}}} \\
\textit{SD v2.1} & Caucasian-Woman & & \cellcolor{Gray}{0.04} & \cellcolor{Gray}{0.24} & \cellcolor{Gray}{\textbf{0.84}} & \redbox{\uarr 3.99} & \bluebox{\darr 4.30} & \bluebox{\darr 4.33} \\
 & Caucasian-Man & & 0 & 0.02 & 0.06 & \redbox{\uarr 3.68} & \redbox{\uarr 1.51} & \redbox{\uarr \textbf{22.52}} \\
 & African-Amer. Woman & & \cellcolor{Gray}{0.02} & \cellcolor{Gray}{0.36} & \cellcolor{Gray}{\textbf{0.78}} & \cellcolor{Gray}{\redbox{\uarr 6.62}} & \cellcolor{Gray}{\bluebox{\darr 0.19}} & \cellcolor{Gray}{\redbox{\uarr \textbf{19.18}}} \\
 & African-Amer. Man & & 0 & 0 & 0.02 & \cellcolor{Gray}{\redbox{\uarr 12.31}} & \cellcolor{Gray}{\bluebox{\darr 1.61}} & \cellcolor{Gray}{\redbox{\uarr \textbf{20.22}}}\\

\midrule
CFD  & White Female & \multirow{2}{5em}{Low-paid professions} & 0.02 & 0.08 & 0.30 &  \bluebox{\darr 0.42} & \bluebox{\darr 4.47} & \bluebox{\darr 8.24}\\
     & White Male  &  & \cellcolor{Gray}{0.38} & \cellcolor{Gray}{\textbf{0.62}} & \cellcolor{Gray}{\textbf{0.56}} & \redbox{\uarr 1.32} & \bluebox{\darr 1.08} & \bluebox{\darr 9.04} \\
     & Black Female & & 0.02 & 0.16 & 0.28 & \cellcolor{Gray}{\redbox{\uarr 1.71}} & \cellcolor{Gray}{\redbox{\uarr 4.62}} & \cellcolor{Gray}{\redbox{\uarr \textbf{24.06}}}\\
     & Black Male & & \cellcolor{Gray}{0.22} & \cellcolor{Gray}{0.42} & \cellcolor{Gray}{\textbf{0.58}} & \cellcolor{Gray}{\redbox{\uarr 0.54}} & \cellcolor{Gray}{\redbox{\uarr 3.05}} & \cellcolor{Gray}{\redbox{\uarr \textbf{20.36}}} \\
\textit{SD v2.1} & Caucasian Woman & & 0.06 & 0.20 & 0.48 & \redbox{\uarr 3.29} & \bluebox{\darr 3.86} & \bluebox{\darr 13.03} \\
 & Caucasian Man & & \cellcolor{Gray}{0.02} & \cellcolor{Gray}{0.20} & \cellcolor{Gray}{0.36} & \redbox{\uarr 8.55} & \redbox{\uarr 7.33} & \redbox{\uarr \textbf{17.05}} \\
 & African-Amer. Woman & & 0.06 & 0.38 & \textbf{0.50} & \cellcolor{Gray}{\redbox{\uarr 10.11}} & \cellcolor{Gray}{\redbox{\uarr 3.69}} & \cellcolor{Gray}{\redbox{\uarr 14.05}}\\
 & African-Amer. Man & & \cellcolor{Gray}{0.22} & \cellcolor{Gray}{0.32} & \cellcolor{Gray}{0.44} & \cellcolor{Gray}{\redbox{\uarr 12.26}} & \cellcolor{Gray}{\redbox{\uarr 8.97}} & \cellcolor{Gray}{\redbox{\uarr \textbf{16.35}}} \\
\bottomrule
\end{NiceTabular}
}
\vspace{-6pt}

\caption{For each row, we edit 25 original images into two professions. The high-paid professions are doctor and CEO. The low-paid professions are \textit{`dishwasher'} and \textit{`fastfood-worker'}. This results in 50 edited images, per edit strength, in each row. `Change in gender (CLIP)' column: Percentage of edited images that are different in gender from original image. We embolden results where more than half the edits alter the gender. `Change in skin-color (ITA)' column: Change in ITA between the average face of the edited set and the original set of images (\bluebox{\darr in ITA} - skin becomes darker, \redbox{\uarr in ITA} - skin becomes lighter). We embolden changes over $\pm$ 15 points. Absolute ITA values are found in Appendix \ref{apx:ITAValues}.}
\vspace{-1.0em}
\label{tab:ResultsTable1}
\end{table*}
\FloatBarrier

\subsection{Results}
\vspace{-2pt}
\subsubsection{Social Bias in Text-guided Image Editing}
\vspace{-2pt}
We varied social groups, target concepts, and edit strengths (refer to Table \ref{tab:ResultsTable1}). 

\textbf{Change in Gender:} Editing images of women towards high-paying careers results in a higher rate of gender alteration than in images of men, at all strengths and in both datasets. This is prominent at the maximum edit strength, 1.0 (81\% vs. 4\% for synthetic and 74\% vs. 8\% for CFD). Editing towards low-paying careers induces a lower rate of gender alteration in images of women and a higher rate in images of men, compared to their respective rates for high-paying careers, at all strengths and in both datasets. This difference is also most prominent at the maximum edit strength (49\% vs. 40\% for synthetic and 57\% vs 29\% for CFD). 

\textbf{Change in Skin-Color:} Positive changes in ITA value between the average face of the original and edited set of images indicate a shift towards lighter tones, while negative changes the opposite. We found an average trend towards skin lightening (M=6.85 for synthetic and M=4.51 for CFD), which is particularly prominent for non-white individuals (M=10.16 for synthetic and M=12.02 for CFD). The shift for non-white individuals is more pronounced at higher edit strengths (0.6: M=10.33, 1.0: M=17.45 for synthetic and 0.6: M=1.13, 1.0: M=29.34 for CFD). High-paid edits have a notably greater increase than low-paid edits at the max edit strength in both datasets (high-paid: M=14.40 , low-paid: M=8.61 for synthetic and high-paid: M=17.48, low-paid: M=6.04 for CFD). These trends are qualitatively validated in the average faces in Appendix \ref{apx:average_faces}.


\subsubsection{Social Bias in Diffusion-based Classification}
\vspace{-2pt}
In Table \ref{tab:ResultsTable2}, we analyze gender bias in a \textit{stable-diffusion-2-1}-based classifier, by measuring association of profession-neutral intersectional female and male image datasets, towards male- and female-dominated profession sets. Association values of 0.0, 0.50, and 1.0 indicate female-only, unbiased, and male-only classifications, respectively. The classifier shows lower-than-neutral association (average across ELBO steps) towards male professions for images of women (0.36 for synthetic and 0.30 for CFD) and higher for images of men (0.65 for synthetic and 0.56 for CFD). The differential association of the male and female datasets intensifies with increased ELBO samples. In the CFD dataset, it changes from 0.16 to 0.36, and in the synthetic dataset, from 0.22 to 0.37, as the ELBO samples is increased from 1 to 100. CFD images of Black females exhibit a less pronounced bias (0.27) at 100 ELBO samples compared to images of White females (0.10), but this interestingly coincides with a much lower gender identification accuracy, 52\% vs. 96\% (see Appendix \ref{apx:gender-classification}).

\vspace{-4pt}
\section{Discussion}
\vspace{-4pt}

\textbf{Social Bias in Diffusion-based Image 
Editing:} 
Frequent unintended alterations to gender and skin color highlights the strong associations between social identities and professions, as reported by prior works in the image generation context \cite{luccioni2023stable, naik2023social}. When the editing prompt challenges prevailing stereotypes associated with the image's identity, ``protected attributes''—characteristics like gender or ethnicity legally safeguarded from discrimination—are frequently modified. The prevalent trend towards lighter skin tones, pronounced in non-white individuals and when editing to high-paying professions, align with prior works that suggest a \emph{``white default''} in image generation models \cite{whitedefault}. The presented biases are acute, as even with visual guidance on protected attributes in the embedding of the original image, edits produce biased and stereotypical results concerning the target prompt. 

\textbf{Social Bias in Diffusion-based Classification:} 
We observe strong differences in the association of neutral images of different genders with particular professions. Increasing the number of ELBO samples improves classification accuracy in \cite{diffusionclassifier, imagenclassifier} and \ref{apx:gender-classification}, but also escalates social bias. Enhanced proficiency in recognizing protected attributes like gender (\ref{apx:gender-classification}) inadvertently intensifies biased associations. This consistent but misleading correlation, in the absence of concrete profession identifiers, raises questions regarding the robustness of the \textit{stable-diffusion-2-1}-based classifiers. 

\textbf{Broader Impact and Deployment:} 
The framework for evaluating the social impact of generative AI systems presented in \cite{solaiman2023evaluating} suggests two modes of evaluation— 1) evaluating the technical 'base' system and 2) the impact of context-specific deployment on people and society. Our work tackles the former, albeit it through hypothetical in-context applications, and raises concerns across several criteria used to evaluate the technical system including 'Bias, Stereotypes, and Representational Harms' and 'Disparate Performance'. Further, the differential ease of applying certain edits to different groups, whilst preserving identity (gender, race but also facial features, for example) makes these systems susceptible to misuse, including for the perpetuation of negative stereotypes. We defer to \cite{solaiman2023evaluating} for further discussion on evaluation areas for potential impacts on people and society.

\textbf{Limitations:} One limitation is the potential for the generator text-to-image model to inject it's own social biases into the test images. We assume generated images are profession neutral and diverse. For attribute concepts where this is not the case, prompt engineering should be explored \cite{bansal2022texttoimage}. We inherit CLIP's fairness and accuracy limitations, by using it for gender classification. However, CLIP achieved a 100\% accuracy on the unedited images, and similar performance is expected for the edited images. Further, image based skin-tone calculation is susceptible to artefacts and low dynamic ranges (note the low initial ITA value for synthetic Caucasian-man images from the occasionally murky generated images). Future work could verify the data quality, by exerting tight control over synthetic image generation (e.g. make attribute level changes to human-curated images) and post-generation normalization (e.g. filter for well-lit images, center faces and normalize viewpoints).

\vspace{-4pt}
\section{Conclusion}
\vspace{-4pt}

\vspace{-4pt}
Methods derived from Stable Diffusion showcased pronounced intersectional biases across gender and skin-color indicating a pressing need for bias testing methods that are aligned with downstream tasks, in order to facilitate ethical deployment. Our work serves as an initial foray into methodology that supports flexible bias testing at scale in two such downstream tasks. In future work, the effectiveness of refined editing or classification prompts \cite{bansal2022texttoimage} and the fairness of diffusion-based classifiers, especially concerning a larger range of intersectional social groups, should be explored to further understand and improve robustness and reliability. 

\begin{ack}
We would like to thank the Caltech SURF program for contributing to the funding of this project. Anima Anandkumar is Bren Professor at Caltech and Senior Director of AI Research at NVIDIA. This material is based upon work supported by the National Science Foundation under Grant \# 2030859 to the Computing Research Association for the CIFellows Project.
\end{ack}

\section{Appendix}

\subsection{Image Editing}
We present the edits for a subset of the edit prompts tested, namely 'doctor' (high-paid) and 'fastfood worker' (low-paid), across both the human-curated (CFD) and synthetic (SD v2.1) sets, for White/Caucasian Men and Black/African-American Women. This provides a qualitative sense of the differential shift of protected attributes, including gender and skin color, when the edit prompt concerns a stereotype or an anti-stereotype concept.

\subsubsection{Human-curated (CFD) Image Edits}
\label{apx:cfd_individual_edits}
\begin{figure}[H]
  \centering
  \includegraphics[width=\linewidth]{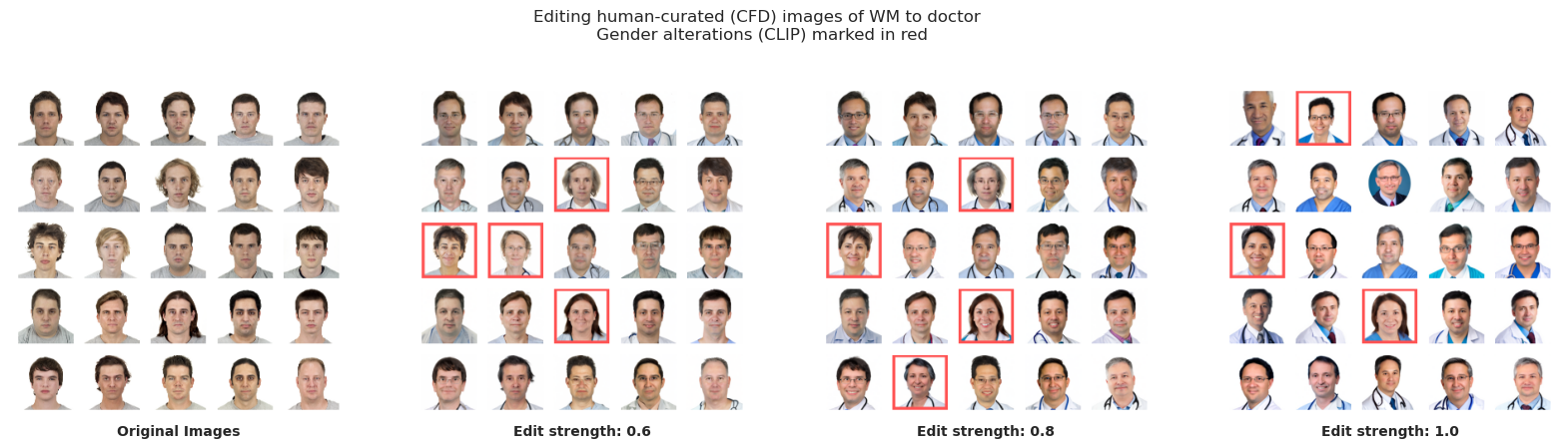}
  \caption{Edits of human-curated (CFD) images of 'White Male' to doctor}
  \label{fig:WM_DOCTOR_TILE}  
\end{figure}
\vspace{-0.25in}
\begin{figure}[H]
  \centering
  \includegraphics[width=\linewidth]{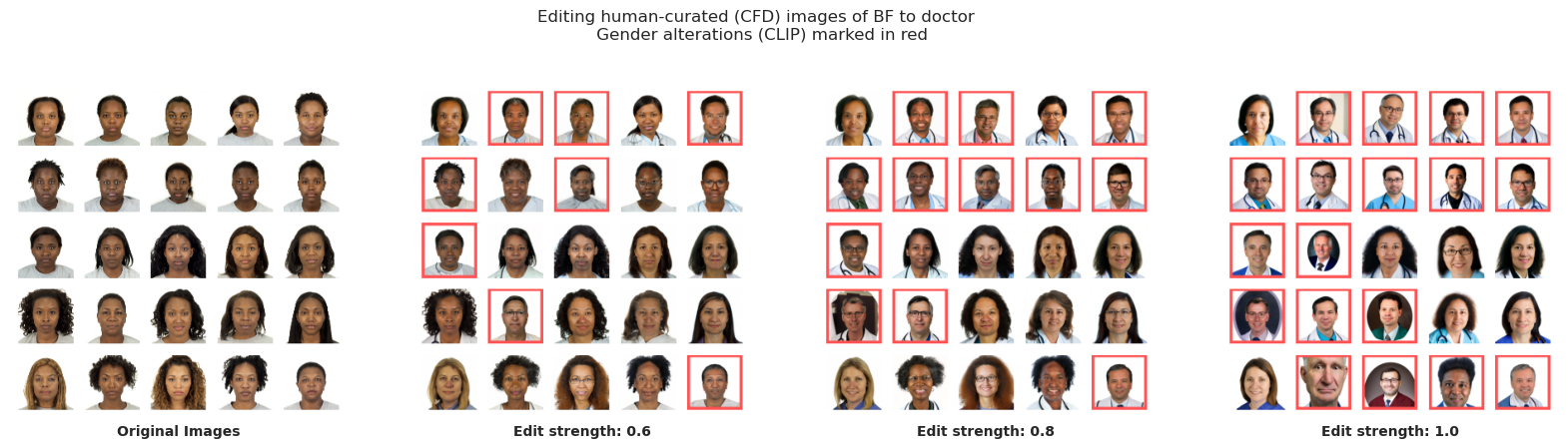}
  \caption{Edits of human-curated (CFD) images of 'Black Female' to doctor}
  \label{fig:BF_DOCTOR_TILE}  
\end{figure}

\begin{figure}[H]
  \centering
  \includegraphics[width=\linewidth]{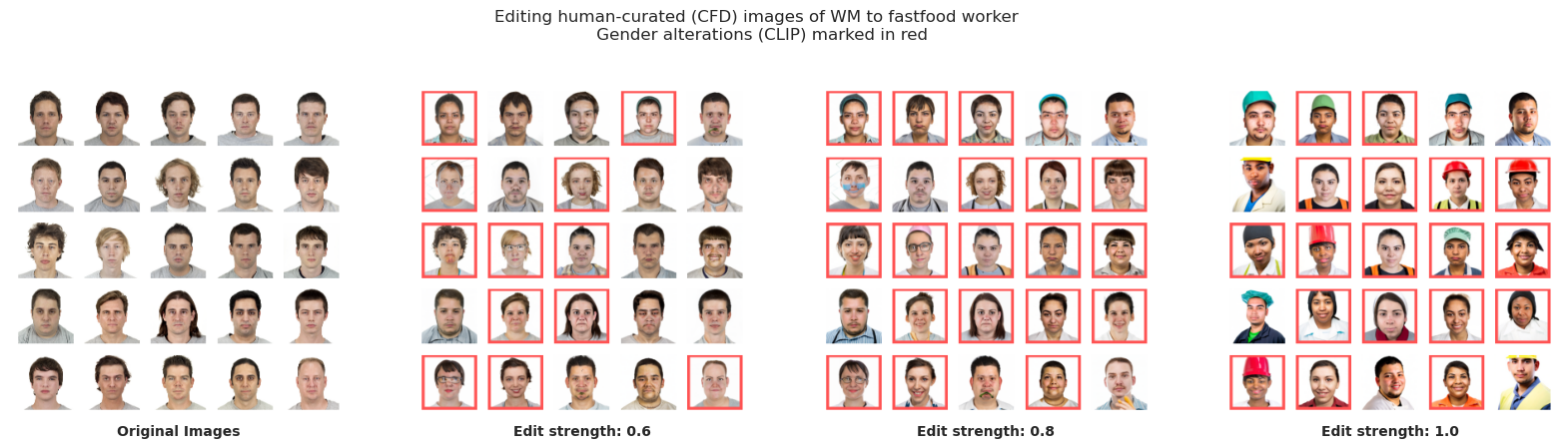}
  \caption{Edits of human-curated (CFD) images of 'White Male' to fastfood worker}
  \label{fig:WM_FASTFOOD_TILE}  
\end{figure}
\vspace{-0.25in}
\begin{figure}[H]
  \centering
  \includegraphics[width=\linewidth]{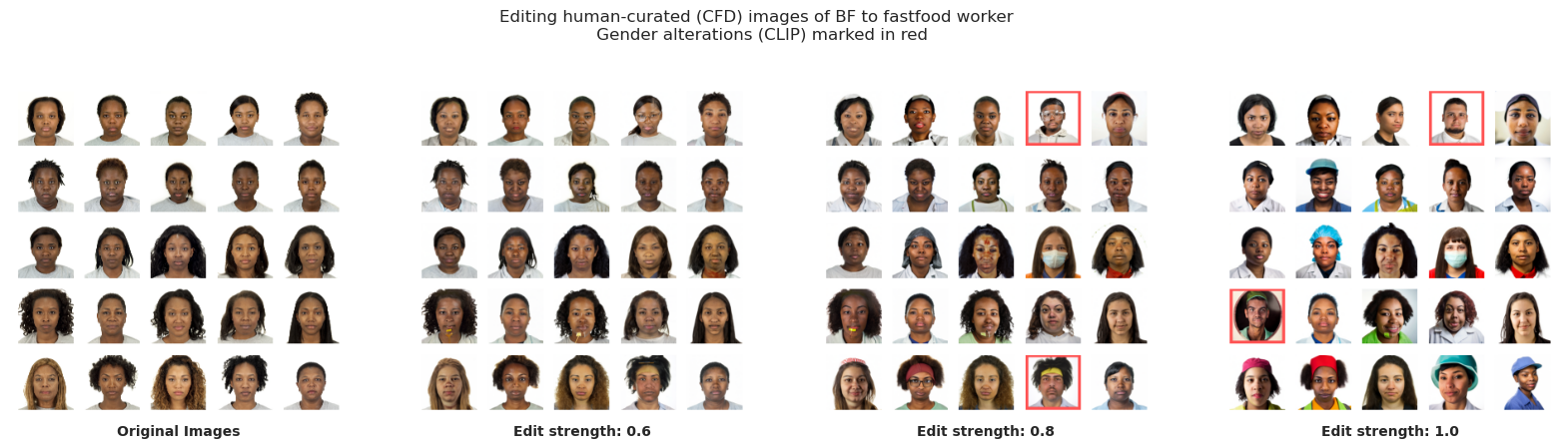}
  \caption{Edits of human-curated (CFD) images of 'Black Female' to fastfood worker}
  \label{fig:BF_FASTFOOD_TILE}  
\end{figure}

\subsubsection{Synthetic (SD v2.1) Image Edits}
\label{apx:sd_individual_edits}

\begin{figure}[H]
  \centering
  \includegraphics[width=\linewidth]{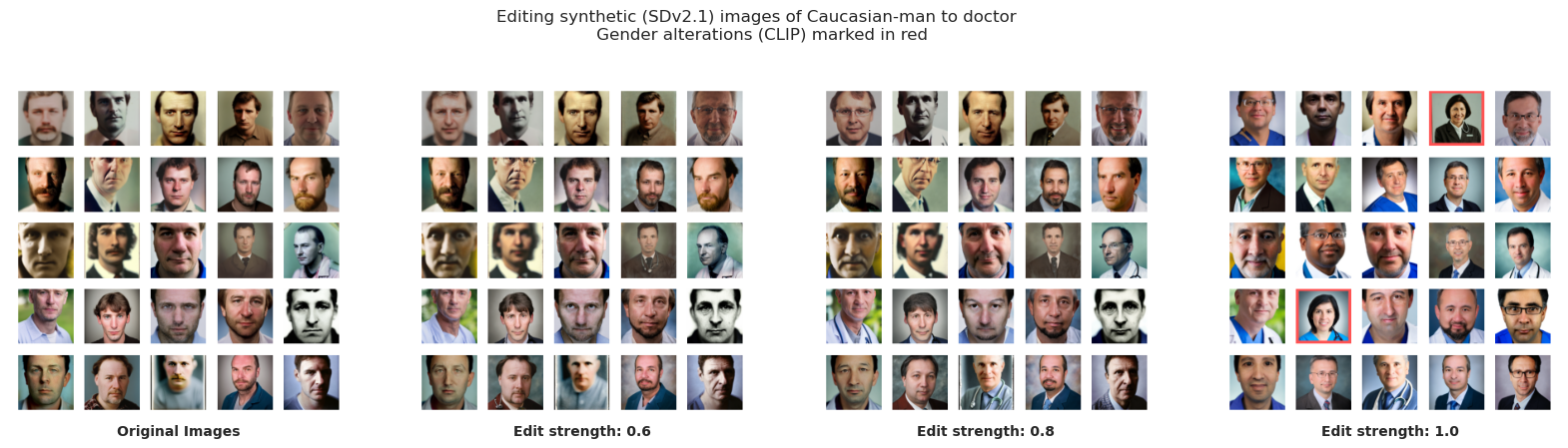}
  \caption{Edits of synthetic (SD v2.1) images of 'Caucasian Man' to doctor}
  \label{fig:Caucasian-Men_DOCTOR_TILE}  
\end{figure}
\vspace{-0.25in}

\begin{figure}[H]
  \centering
  \includegraphics[width=\linewidth]{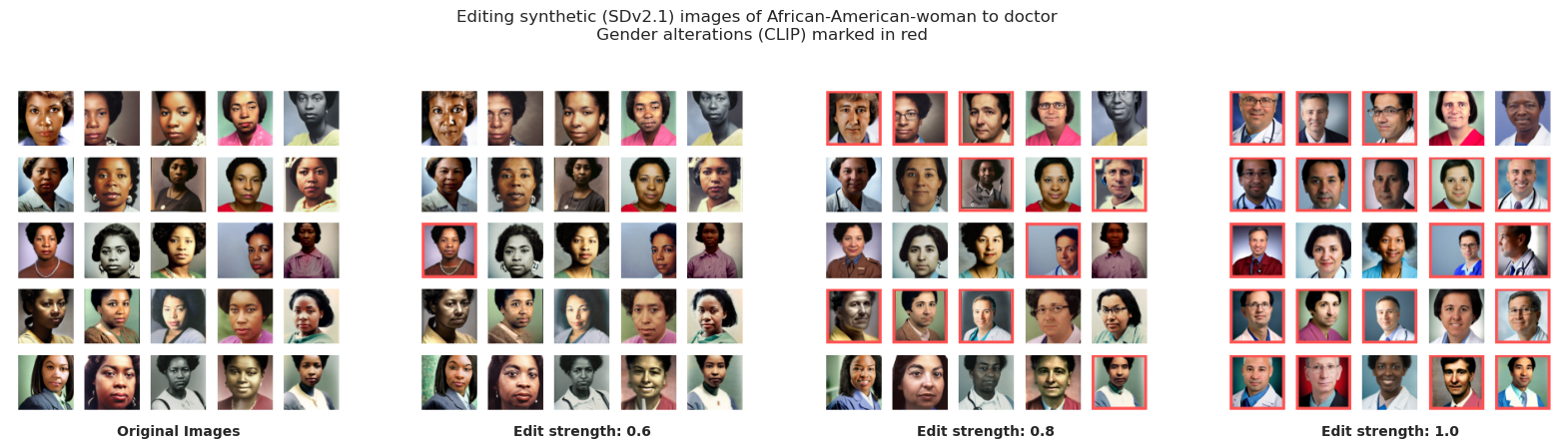}
  \caption{Edits of synthetic (SD v2.1) images of 'African-American Woman' to doctor}
  \label{fig:African-American-Woman_DOCTOR_TILE}  
\end{figure}
\vspace{-0.25in}

\begin{figure}[H]
  \centering
  \includegraphics[width=\linewidth]{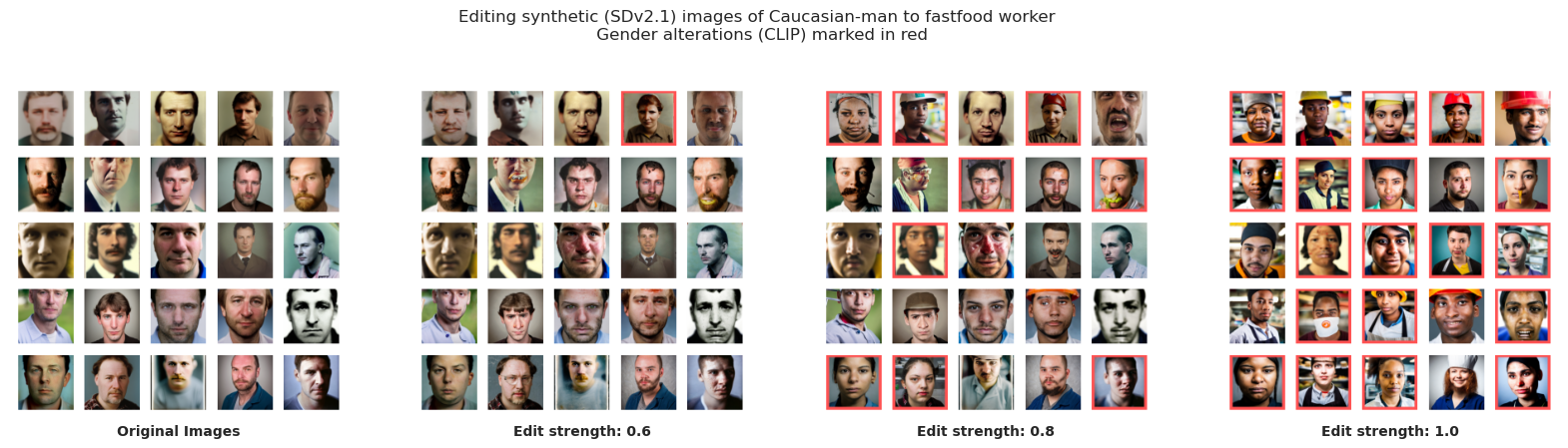}
  \caption{Edits of synthetic (SD v2.1) images of 'Caucasian Man' to fastfood worker}
  \label{fig:Caucasian-Men_FASTFOOD_TILE}  
\end{figure}
\vspace{-0.25in}

\begin{figure}[H]
  \centering
  \includegraphics[width=\linewidth]{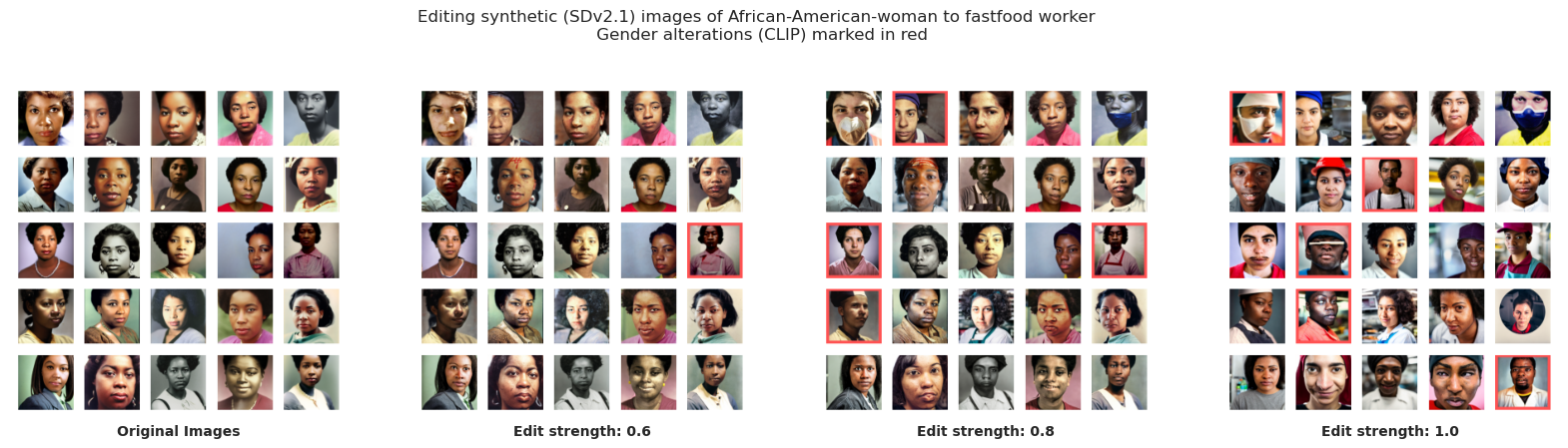}
  \caption{Edits of synthetic (SD v2.1) images of 'African-American Woman' to fastfood worker}
  \label{fig:African-American-Woman_FASTFOOD_TILE}  
\end{figure}
\vspace{-0.25in}

\subsubsection{Average Faces}
\label{apx:average_faces}
We present the average faces of the original images, of those edited, and the edits, of different edit strengths and concepts, for all social groups, for both human-curated (CFD) and synthetic data.

\begin{figure}[htbp]
  \centering
  \includegraphics[width=\linewidth]{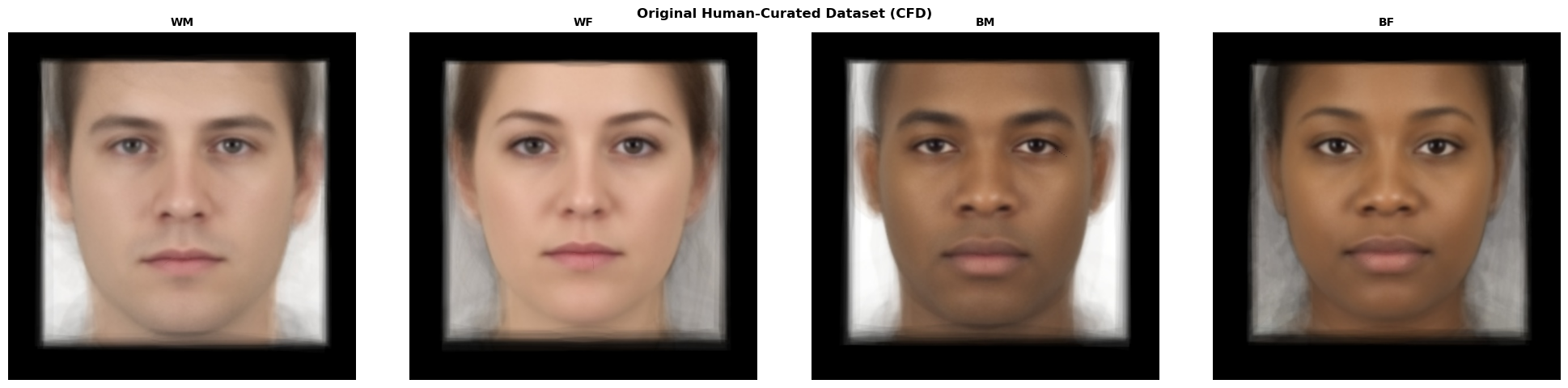}
  \caption{Average faces of the images that were edited (Original) - Human-curated (CFD)}
  \label{fig:average_faces_cfd_original}  
\end{figure}

\begin{figure}[htbp]
  \centering
  \includegraphics[width=\linewidth]{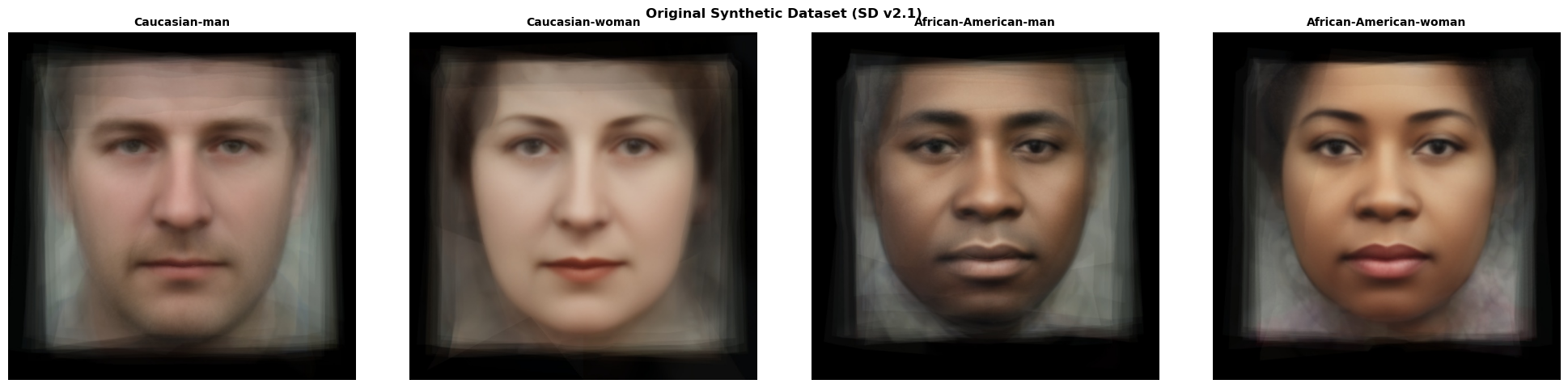}
  \caption{Average faces of the images that were edited (Original) - Synthetic (SD v2.1)}
  \label{fig:average_faces_syn_original}  
\end{figure}

\begin{figure}[htbp]
  \centering
  \includegraphics[width=\linewidth]{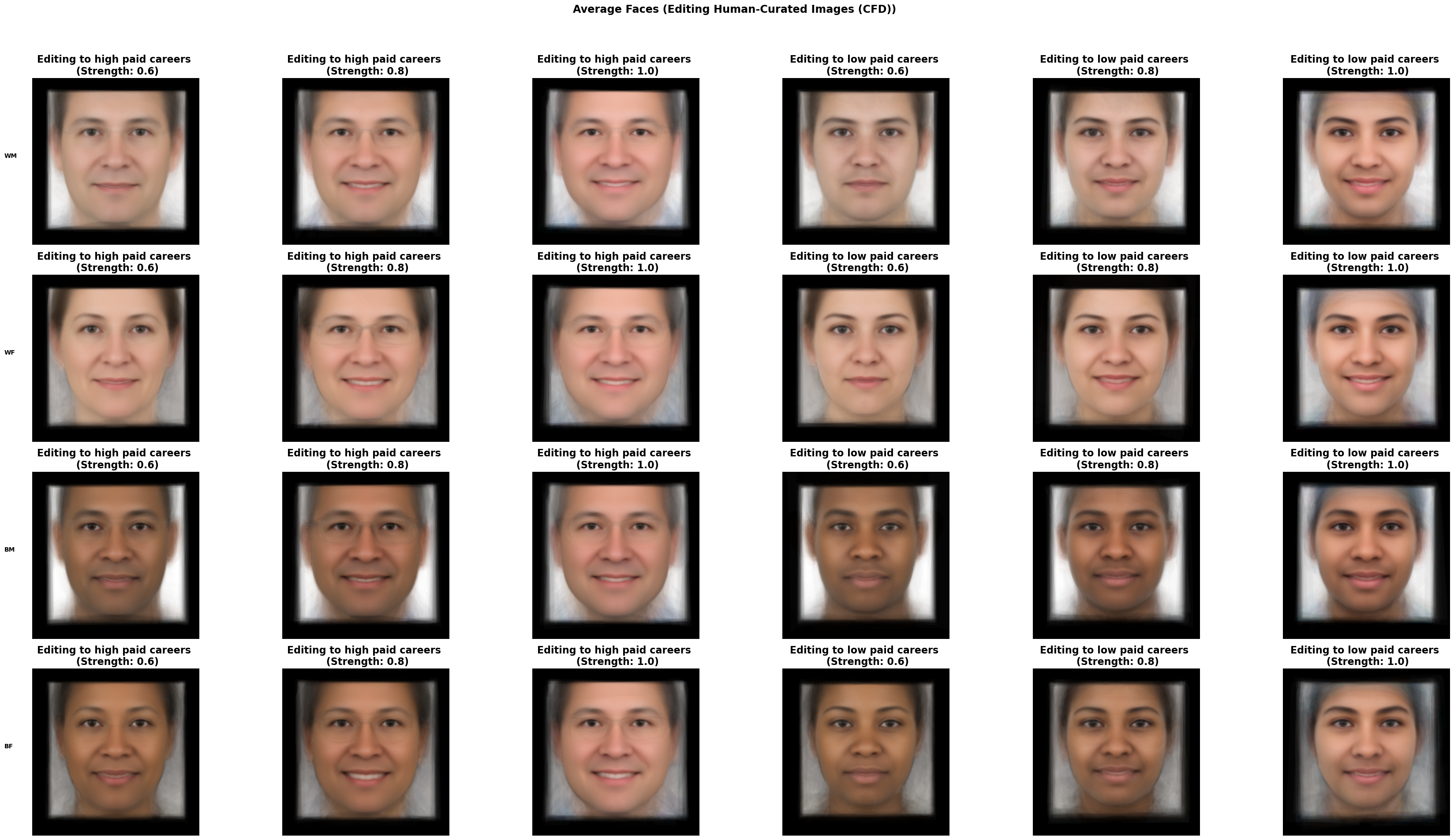}
  \caption{Average faces of edited sets of images, for human-curated (CFD) images, when edited towards high- and low-paying careers. Note that each average face is comprised of 50 images, 25 edits towards each profession (low-paying: dishwasher-worker and fastfood-worker, high-paying: doctor and CEO).}
  \label{fig:average_faces_cfd_edits}  
\end{figure}

\begin{figure}[htbp]
  \centering
  \includegraphics[width=\linewidth]{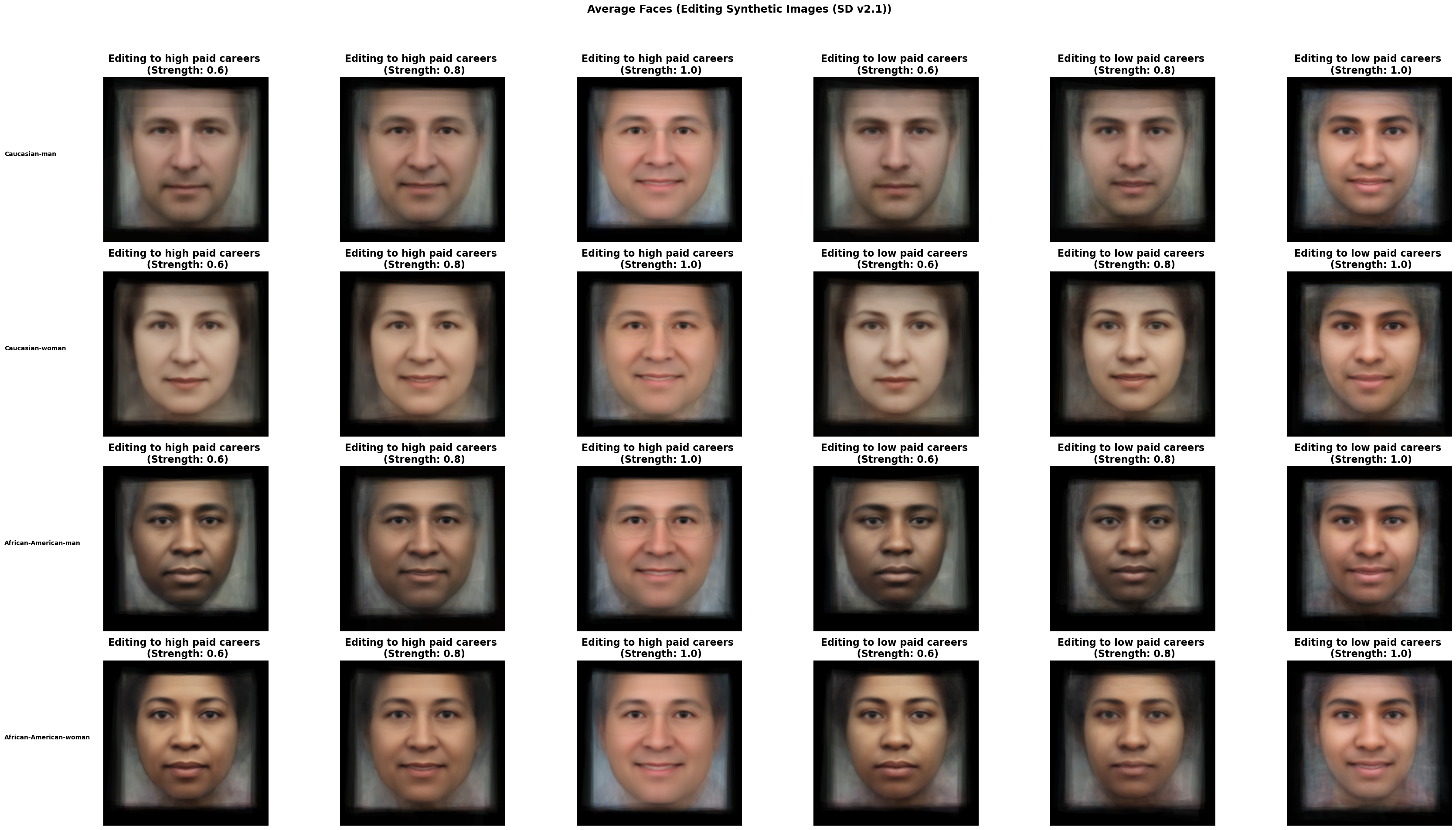}
  \caption{Average faces of edited sets of images, for synthetic (SD v2.1) images, when edited towards high- and low-paying careers. Note that each average face is comprised of 50 images, 25 edits towards each profession (low-paying: dishwasher-worker and fastfood-worker, high-paying: doctor and CEO).}
  \label{fig:average_faces_syn_edits}  
\end{figure}

\FloatBarrier

\subsubsection{Average Faces - Skin-color (ITA) Values}
\label{apx:ITAValues}
\small{
\begin{table}[htpb]
    \centering
    \begin{tabular}{@{}l l r@{}}
        \textbf{Dataset} & \textbf{Social group} & \textbf{ITA} \\ 
\midrule
        CFD & Black Male & -17.73 \\ 
        & White Male & 41.91 \\ 
        & Black Female & -15.66 \\ 
        & White Female & 39.39 \\ 
        \textit{SD v2.1} & Caucasian-woman & 26.63 \\ 
        & Caucasian-man & 2.52 \\ 
        & African-American-man & -10.29 \\ 
        & African-American-woman & -3.81 \\
\bottomrule
    \end{tabular}"
    \vspace{1.0em}
\caption{We report the absolute ITA values for the average face of the original images that were edited, for each social group, for both human-curated (CFD) and synthetic (\textit{SD v2.1}) datasets (averages faces shown in \ref{fig:average_faces_cfd_original} and \ref{fig:average_faces_syn_original}).}
\vspace{-2.0em}
\label{tab:ITATableOriginal}
\end{table}

\begin{table*}[!htbp]
    \centering
    \small{
    \begin{tabular}{@{}l l l l l l@{}}
        \textbf{Dataset} & \textbf{Social group} & \textbf{Edit Concept} & ITA (0.6) & ITA (0.8) & ITA (1.0) \\ 
\midrule
        CFD & BM & low-paid & -17.18 & -14.68 & 2.63 \\ 
        ~ &  & high-paid & -17.16 & -11.64 & 17.84 \\ 
        ~ & BF & low-paid & -13.94 & -11.03 & 8.4 \\ 
        ~ &  & high-paid & -13.94 & -7.06 & 21.74 \\ 
        ~ & WM & low-paid & 43.23 & 40.83 & 32.87 \\ 
        ~ &  & high-paid & 40.76 & 35.87 & 40.51 \\ 
        ~ & WF & low-paid & 38.97 & 34.91 & 31.15 \\ 
        ~ &  & high-paid & 39.71 & 35.14 & 37.74 \\ 
        \textit{SD v2.1} & African-American-man & low-paid & 1.97 & -1.32 & 6.06 \\
        ~ &  & high-paid & 2.02 & -11.9 & 9.93 \\ 
        ~ & African-American-woman & low-paid & 6.29 & -0.12 & 10.24 \\ 
        ~ &  & high-paid & 2.81 & -4.01 & 15.36 \\ 
        ~ & Caucasian-man & low-paid & 11.07 & 9.85 & 19.57 \\ 
        ~ & & high-paid & 6.2 & 4.03 & 25.04 \\ 
        ~ & Caucasian-woman & low-paid & 29.93 & 22.78 & 13.61 \\ 
        ~ & & high-paid & 30.62 & 22.33 & 22.31 \\ 
\bottomrule
    \end{tabular}}
    \vspace{1.0em}
\caption{We report the absolute ITA values for the average face of the edited sets of images, for both human-curated (CFD) and synthetic (\textit{SD v2.1}) datasets, at all edit strengths, for low and high-paid professions (average faces shown in \ref{fig:average_faces_cfd_edits} and \ref{fig:average_faces_syn_edits}).}
\vspace{-2.0em}
\label{tab:ITATableEdits}
\end{table*}

\subsection{Classification}

\subsubsection{Gender Classification}
\label{apx:gender-classification}

\begin{table*}[htbp]
\centering
\small{
\begin{tabular}{@{}l l r r r @{}}
\textbf{Dataset} & \textbf{Social group} & \multicolumn{3}{c}{\textbf{Accuracy}}\\
\toprule
\multicolumn{1}{c}{Number of noise samples in ELBO estimation} & & 1 & 10 & 100\\
\midrule
CFD & White Female & 0.54 & 0.65 & 0.96 \\
& White Male & 0.75  & 0.93 & 0.95 \\
& Black Female & 0.35 & 0.35  & 0.52 \\
& Black Male & 0.93 & 0.98 & 1.00 \\
\textit{SD v2.1} & Caucasian-Woman & 0.68 & 0.87 & 0.96 \\
& Caucasian-Man & 0.83 & 0.86 & 0.97 \\
& African-American-Woman & 0.71 & 0.87 & 0.97\\
& African-American-Man & 0.69 & 0.80 & 0.97 \\
\bottomrule
\end{tabular}}
\vspace{1.0em}
\caption{We report accuracy in gender classification, into 'A portrait of a man.' and 'A portrait of a woman.', as a reference of the classification fidelity of the SD v2.1-based classifier. We report results for 1, 10 and 100 noise samples in the estimation of the classification objective (ELBO). Accuracy increases across the board as the number of the noise samples is increased.}
\vspace{-2.0em}
\label{tab:ResultsTable3}
\end{table*}
\FloatBarrier

\clearpage
\subsubsection{Profession Associations with Intersectional Social Identities}

\begin{table*}[htbp]
\centering
\small{
\begin{NiceTabular}{@{}l l l l r r r @{}}

\textbf{Dataset} & {\bf Social Identity ($X$)} & Target Set 1 (\textbf{$A$}) & Target Set 2 (\textbf{$B$}) & \multicolumn{3}{c}{\textbf{$S(\bm{\mathcal{D}_{X}}, \bm{A}, \bm{B})$}}\\
\toprule
\multicolumn{4}{l}{Number of noise samples in ELBO estimation}  & 1 & 10 & 100\\
\midrule
CFD & White Female & Male-dominated & Female-dominated & 0.42 & 0.18 & 0.10 \\
\rowcolor{Gray}
\cellcolor{White}{} & White Male & professions & professions & {\bf 0.54} & 0.47 & {\bf 0.51} \\
 & Black Female &  & & 0.47 & 0.39 & 0.27\\
 \rowcolor{Gray}
\cellcolor{White}{} & Black Male &  &   & {\bf 0.67} & {\bf 0.62} & {\bf 0.58} \\
\textit{SD v2.1} & Caucasian Woman &  &  & 0.39 & 0.30 & 0.35\\
\rowcolor{Gray}
\cellcolor{White}{} & Caucasian Man &  &   & {\bf 0.67} & {\bf 0.66} & {\bf 0.83} \\
 & African-Amer. Woman &  & & 0.43 & 0.33 & 0.39 \\
\rowcolor{Gray}
\cellcolor{White}{} & African-Amer. Man &  & & {\bf 0.59} & {\bf 0.51} & {\bf 0.65}\\
\midrule
\multirow{2}{5em}{Mean across races} & {\bf Female} & & & 0.43 & 0.30 & 0.28 \\
 & \cellcolor{Gray}{\bf Male} & \cellcolor{Gray}{} & \cellcolor{Gray}{} & \cellcolor{Gray}{\bf 0.62} & \cellcolor{Gray}{\bf 0.57} & \cellcolor{Gray}{\bf 0.64} \\
\bottomrule
\end{NiceTabular}}

\vspace{-6pt}
\caption{Association measure towards male- and female-dominated profession sets, for human-curated (CFD) and synthetic (SD v2.1) datasets across inter-sectional social identities. We embolden values greater than 0.50, which suggests a biased association towards the male-dominated professions set, $A$.
}
\vspace{-1.0em}
\label{tab:ResultsTable2}
\end{table*}

\FloatBarrier

\clearpage
\subsubsection{Profession Associations - Individual Comparisons}\label{apx:biased_classification}
We present the inter target-set comparisons from the CFD and synthetic images' association tests in which one of the two profession classes is picked > 75\% of the time. We report the results for 100 ELBO samples, the set-up that corresponds to the greatest classifier fidelity of those tested (1, 10, 100).

\begin{table}[htbp]
    \centering
    \small{
    \begin{tabular}{llll}
        $X$ & a & b & \%a \\ 
        \toprule
        BF & mechanic & babysitter & 0.0 \\ 
        BF & mechanic & receptionist & 0.0 \\ 
        BF & mechanic & housekeeper & 0.09 \\ 
        BF & mechanic & nurse & 0.16 \\ 
        BF & plumber & babysitter & 0.0 \\ 
        BF & plumber & receptionist & 0.01 \\ 
        BF & plumber & housekeeper & 0.2 \\ 
        BF & carpenter & babysitter & 0.0 \\ 
        BF & carpenter & secretary & 0.77 \\ 
        BF & carpenter & receptionist & 0.01 \\ 
        BF & carpenter & housekeeper & 0.24 \\ 
        BF & construction worker & babysitter & 0.0 \\ 
        BF & construction worker & secretary & 0.83 \\ 
        BF & construction worker & receptionist & 0.04 \\ 
        BF & truck driver & babysitter & 0.0 \\ 
        BF & truck driver & secretary & 0.77 \\ 
        BF & truck driver & receptionist & 0.04 \\ 
        BM & mechanic & babysitter & 0.06 \\ 
        BM & mechanic & secretary & 1.0 \\ 
        BM & mechanic & receptionist & 0.12 \\ 
        BM & mechanic & housekeeper & 0.77 \\ 
        BM & mechanic & nurse & 0.79 \\ 
        BM & plumber & babysitter & 0.1 \\ 
        BM & plumber & secretary & 1.0 \\ 
        BM & plumber & receptionist & 0.16 \\ 
        BM & plumber & housekeeper & 0.87 \\ 
        BM & plumber & nurse & 0.81 \\ 
        BM & carpenter & babysitter & 0.05 \\ 
        BM & carpenter & secretary & 1.0 \\ 
        BM & carpenter & receptionist & 0.13 \\ 
        BM & carpenter & housekeeper & 0.83 \\ 
        BM & carpenter & nurse & 0.75 \\ 
        BM & construction worker & babysitter & 0.14 \\ 
        BM & construction worker & secretary & 0.99 \\ 
        BM & construction worker & receptionist & 0.21 \\ 
        BM & construction worker & housekeeper & 0.85 \\ 
        BM & construction worker & nurse & 0.78 \\ 
        BM & truck driver & babysitter & 0.16 \\ 
        BM & truck driver & secretary & 1.0 \\ 
        BM & truck driver & receptionist & 0.24 \\ 
        \bottomrule
        \end{tabular}
        \hfill
        \begin{tabular}{llll}
            $X$ & a & b & \%a\\ 
            \toprule
        BM & truck driver & housekeeper & 0.87 \\ 
        BM & truck driver & nurse & 0.84 \\ 
        WF & mechanic & babysitter & 0.0 \\ 
        WF & mechanic & secretary & 0.21 \\ 
        WF & mechanic & receptionist & 0.0 \\ 
        WF & mechanic & housekeeper & 0.03 \\ 
        WF & mechanic & nurse & 0.0 \\ 
        WF & plumber & babysitter & 0.0 \\ 
        WF & plumber & receptionist & 0.0 \\ 
        WF & plumber & housekeeper & 0.06 \\ 
        WF & plumber & nurse & 0.21 \\ 
        WF & carpenter & babysitter & 0.0 \\ 
        WF & carpenter & receptionist & 0.0 \\ 
        WF & carpenter & housekeeper & 0.01 \\ 
        WF & carpenter & nurse & 0.03 \\ 
        WF & construction worker & babysitter & 0.0 \\ 
        WF & construction worker & receptionist & 0.01 \\ 
        WF & construction worker & housekeeper & 0.04 \\ 
        WF & construction worker & nurse & 0.14 \\ 
        WF & truck driver & babysitter & 0.0 \\ 
        WF & truck driver & receptionist & 0.01 \\ 
        WF & truck driver & housekeeper & 0.05 \\ 
        WF & truck driver & nurse & 0.08 \\ 
        WM & mechanic & babysitter & 0.0 \\ 
        WM & mechanic & secretary & 0.93 \\ 
        WM & mechanic & receptionist & 0.13 \\ 
        WM & plumber & babysitter & 0.17 \\ 
        WM & plumber & secretary & 0.96 \\ 
        WM & plumber & housekeeper & 0.78 \\ 
        WM & plumber & nurse & 0.83 \\ 
        WM & carpenter & babysitter & 0.02 \\ 
        WM & carpenter & secretary & 0.96 \\ 
        WM & carpenter & receptionist & 0.18 \\ 
        WM & construction worker & babysitter & 0.04 \\ 
        WM & construction worker & secretary & 0.97 \\ 
        WM & construction worker & receptionist & 0.21 \\ 
        WM & truck driver & babysitter & 0.08 \\ 
        WM & truck driver & secretary & 0.97 \\ 
        WM & truck driver & nurse & 0.81 \\ 
        \bottomrule
        \end{tabular}}
\vspace{1.0em}
\caption{Inter target-set comparisons with >75\% decisions towards one profession for human-curated (CFD) images (100 ELBO samples)}
\label{tab:SignificantClassificationBiasCFD}
\end{table}
\FloatBarrier

\begin{table}[t]
\centering
\small{
    \centering
    \begin{tabular}{llll}
        $X$ & a & b & \%a \\
        \toprule
        African-American-man & mechanic & babysitter & 0.87 \\ 
        African-American-man & mechanic & secretary & 0.85 \\ 
        African-American-man & mechanic & housekeeper & 0.8 \\ 
        African-American-man & mechanic & nurse & 0.24 \\ 
        African-American-man & plumber & receptionist & 0.15 \\ 
        African-American-man & plumber & nurse & 0.12 \\ 
        African-American-man & carpenter & babysitter & 0.94 \\ 
        African-American-man & carpenter & secretary & 0.93 \\ 
        African-American-man & carpenter & housekeeper & 0.85 \\ 
        African-American-man & construction worker & babysitter & 0.82 \\ 
        African-American-man & construction worker & secretary & 0.83 \\ 
        African-American-man & construction worker & housekeeper & 0.82 \\ 
        African-American-man & truck driver & babysitter & 0.94 \\ 
        African-American-man & truck driver & secretary & 0.88 \\ 
        African-American-man & truck driver & receptionist & 0.82 \\ 
        African-American-man & truck driver & housekeeper & 0.88 \\ 
        African-American-woman & mechanic & receptionist & 0.24 \\ 
        African-American-woman & mechanic & nurse & 0.01 \\ 
        African-American-woman & plumber & babysitter & 0.02 \\ 
        African-American-woman & plumber & secretary & 0.15 \\ 
        African-American-woman & plumber & receptionist & 0.01 \\ 
        African-American-woman & plumber & nurse & 0.0 \\ 
        African-American-woman & carpenter & secretary & 0.81 \\ 
        African-American-woman & carpenter & housekeeper & 0.77 \\ 
        African-American-woman & carpenter & nurse & 0.05 \\ 
        African-American-woman & construction worker & nurse & 0.05 \\ 
        African-American-woman & truck driver & nurse & 0.12 \\ 
        Caucasian-man & mechanic & babysitter & 0.94 \\ 
        Caucasian-man & mechanic & secretary & 0.95 \\ 
        Caucasian-man & mechanic & receptionist & 0.96 \\ 
        Caucasian-man & mechanic & housekeeper & 0.86 \\ 
        Caucasian-man & plumber & babysitter & 0.75 \\ 
        Caucasian-man & plumber & secretary & 0.86 \\ 
        Caucasian-man & plumber & receptionist & 0.78 \\ 
        Caucasian-man & plumber & housekeeper & 0.79 \\ 
        Caucasian-man & carpenter & babysitter & 0.95 \\ 
        Caucasian-man & carpenter & secretary & 0.99 \\ 
        Caucasian-man & carpenter & receptionist & 0.98 \\ 
        Caucasian-man & carpenter & housekeeper & 0.91 \\ 
        Caucasian-man & construction worker & babysitter & 0.86 \\ 
        Caucasian-man & construction worker & secretary & 0.92 \\ 
        Caucasian-man & construction worker & receptionist & 0.93 \\ 
        Caucasian-man & construction worker & housekeeper & 0.84 \\ 
        Caucasian-man & truck driver & babysitter & 0.96 \\ 
        Caucasian-man & truck driver & secretary & 0.96 \\ 
        Caucasian-man & truck driver & receptionist & 0.99 \\ 
        Caucasian-man & truck driver & housekeeper & 0.9 \\ 
        Caucasian-woman & mechanic & nurse & 0.01 \\ 
        Caucasian-woman & plumber & babysitter & 0.03 \\ 
        Caucasian-woman & plumber & secretary & 0.12 \\ 
        Caucasian-woman & plumber & receptionist & 0.06 \\ 
        Caucasian-woman & plumber & nurse & 0.0 \\ 
        Caucasian-woman & carpenter & nurse & 0.03 \\ 
        Caucasian-woman & construction worker & nurse & 0.02 \\ 
        Caucasian-woman & truck driver & nurse & 0.05 \\ 
    \bottomrule
    \end{tabular}}
    \caption{Inter target-set comparisons with >75\% decisions towards one profession for synthetic (SD v2.1) images (100 ELBO samples)}
\label{tab:SignificantClassificationBiasSYN}
\end{table}
\FloatBarrier

\subsubsection{Visualisation}\label{apx:biased_classification_visualisation}
A significant inter target-set comparison/classification task, between classes a: construction-worker and b: babysitter, from the synthetic images' (SD v2.1) association test is visualised below. This provides a qualitative sense of the spread of professions assigned to images of different social identities. The figure is generated from classification results in which 100 samples were used in ELBO estimation.

\begin{figure}[htbp]
  \centering
  \begin{minipage}{0.48\linewidth}
      \includegraphics[width=\linewidth]{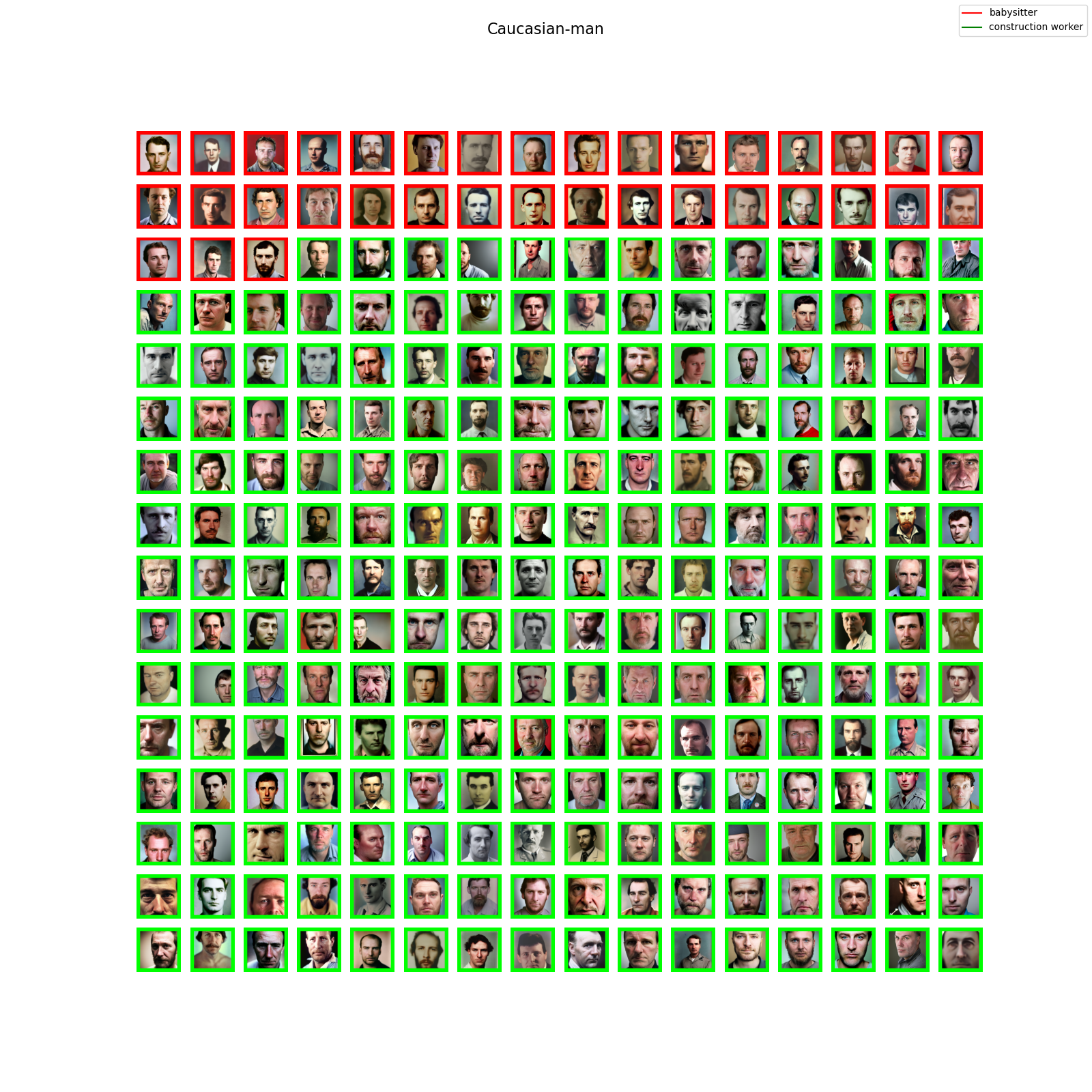}
      \caption{Synthetic (SD v2.1) `Caucasian Man' images classified into classes `babysitter' and `construction worker'}
      \label{CMBabysitterConstructionworker}
  \end{minipage}
  \hfill
  \begin{minipage}{0.48\linewidth}
        \includegraphics[width=\linewidth]{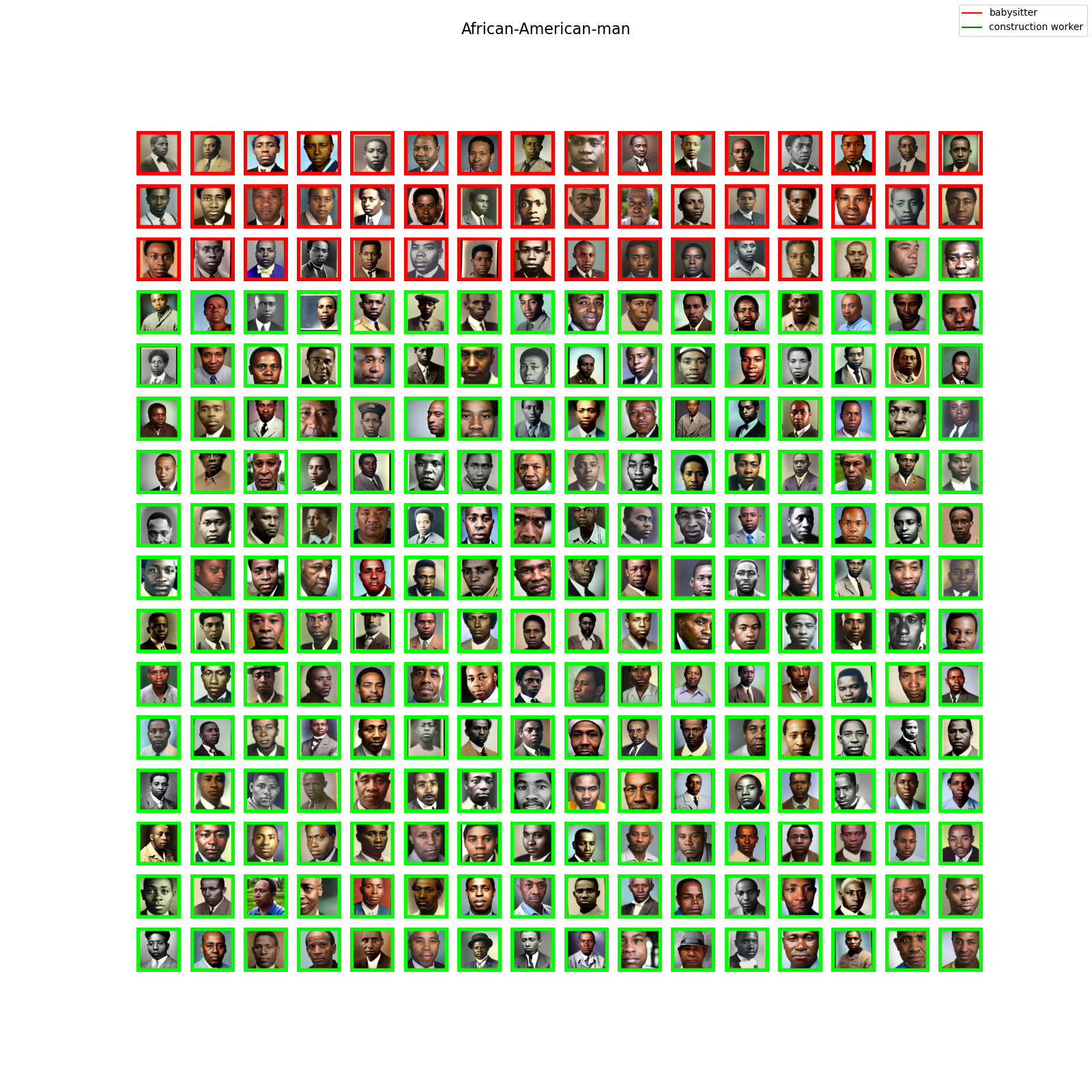}
      \caption{Synthetic  (SD v2.1) 'African-American man' images classified into classes 'babysitter' and 'construction worker'}
          \label{AFMBabysitterConstructionworker}
   \end{minipage}
\end{figure}

\begin{figure}[htbp]
  \centering
  \begin{minipage}{0.48\linewidth}
      \includegraphics[width=\linewidth]{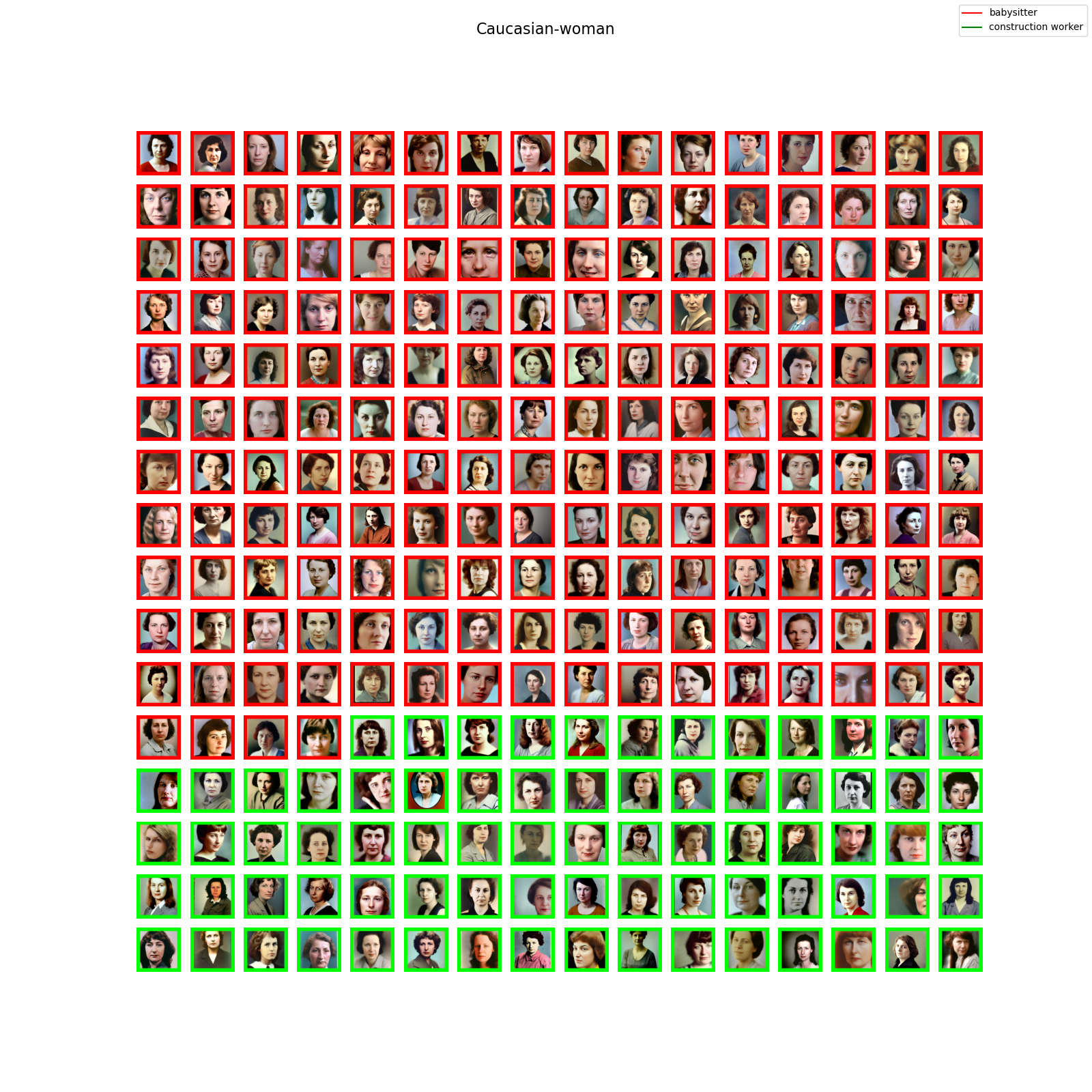}
      \caption{Synthetic (SD v2.1) `Caucasian Woman' images classified into classes `babysitter' and `construction worker'}
      \label{CWBabysitterConstructionworker}
  \end{minipage}
  \hfill
  \begin{minipage}{0.48\linewidth}
        \includegraphics[width=\linewidth]{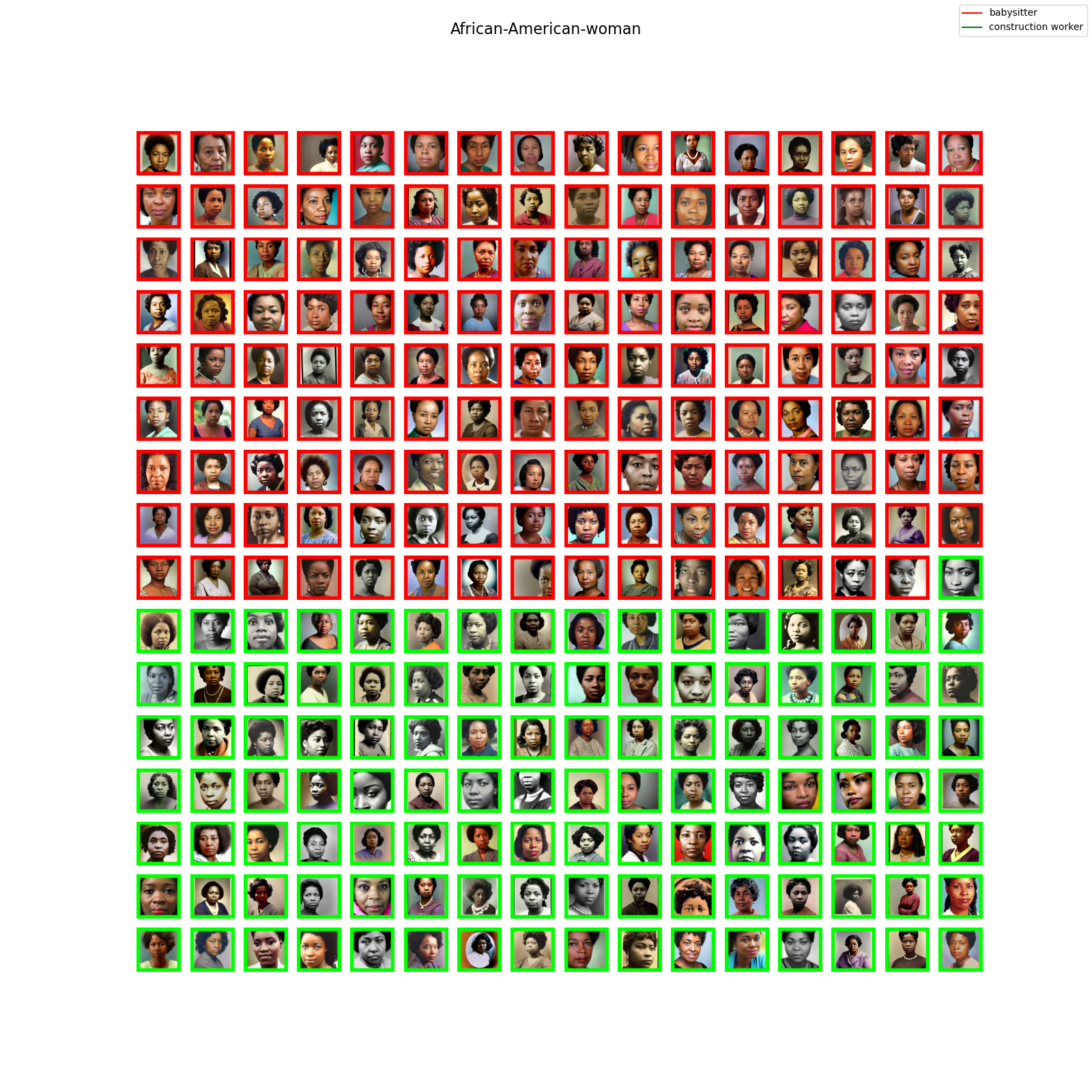}
      \caption{Synthetic (SD v2.1) 'African-American woman' images classified into classes 'babysitter' and 'construction worker'}
          \label{AFWMBabysitterConstructionworker}
   \end{minipage}
\end{figure}

\clearpage
\subsection{Aggregated Bias Results - Human-curated (CFD) images}\label{apx:human_aggregated_bias_results}
\begin{figure}[htbp]
  \centering
  \includegraphics[width=\linewidth]{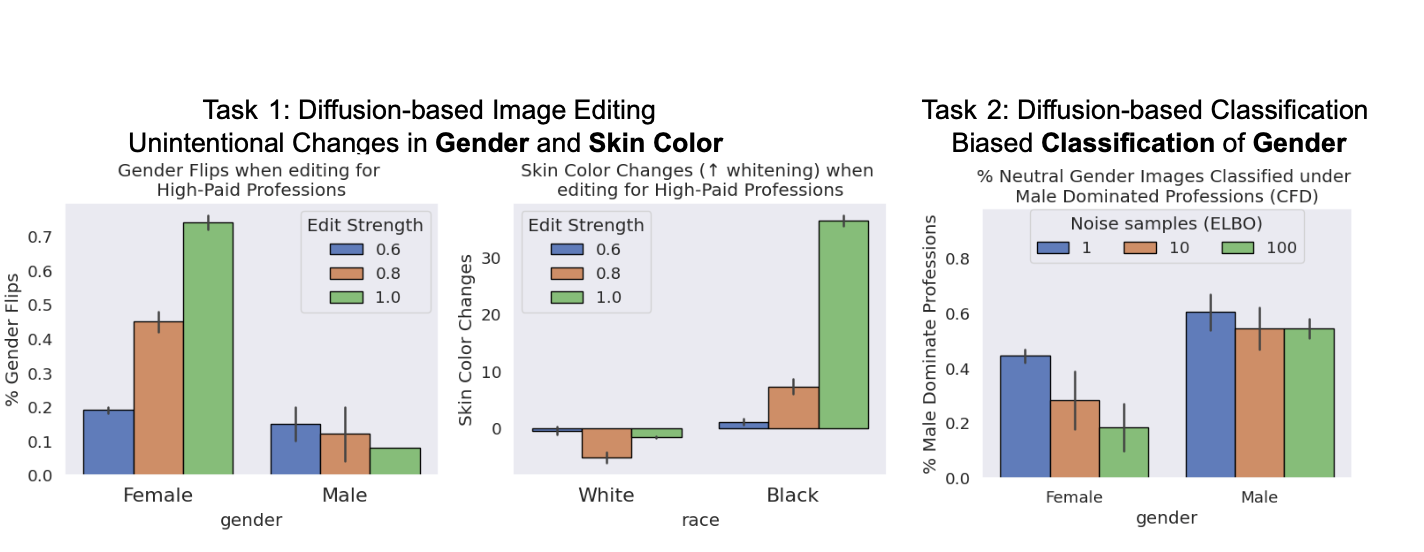}
  \vspace{-12.0pt}
  \caption{Left: Percentage of flips in gender (CLIP) from editing human-curated (CFD) Male and Female images using high-paid prompts in diffusion-based image editing. Middle: Skin Color Changes ($\uparrow$ change towards lighter skin color using an established methodology described in \Sref{sec:methodlogy}) from editing human-curated (CFD) images of White and Black individuals using high-paid prompts in diffusion-based image editing. Right: Percentage of diffusion-based classifier choices towards male-dominated professions in binary classification tasks between a male- and female-dominated profession pair (at different numbers of noise samples in the estimation of the classification objective) for human-curated (CFD) images.}
  \vspace{-9.0pt}
  \label{fig:BiasDownstreamGraph-CFD}  
\end{figure}

\subsection{Aggregated Bias Results - Synthetic (SD) images}\label{apx:synthetic_aggregated_bias_results}

\begin{figure}[htbp]
  \centering
  \includegraphics[width=\linewidth]{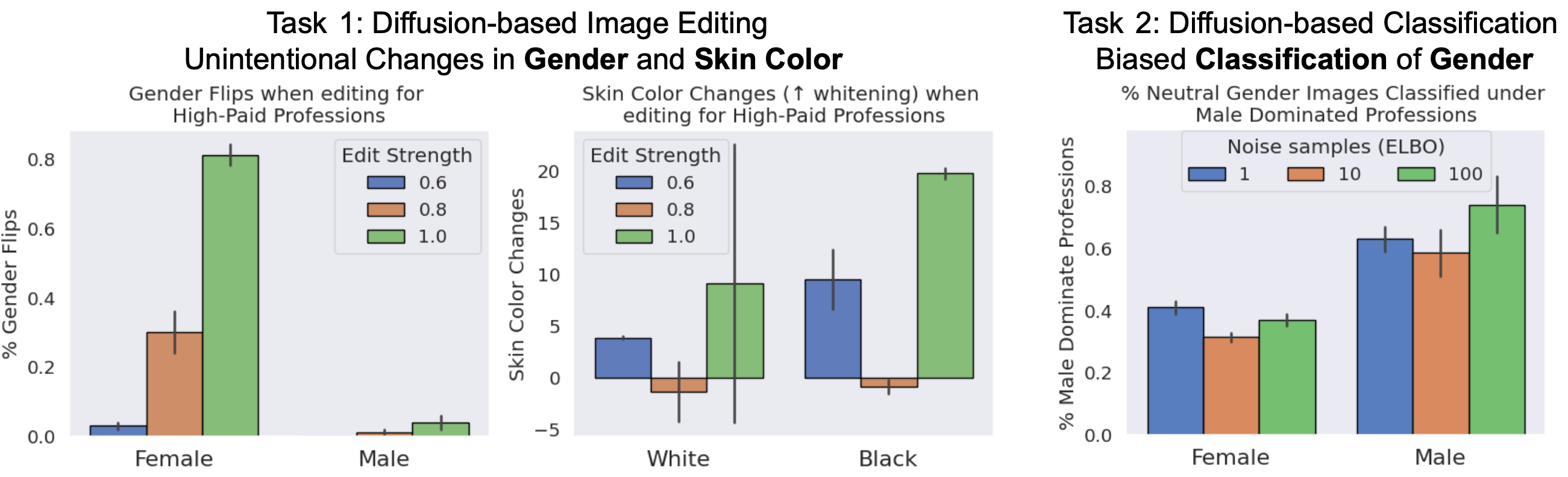}
  \vspace{-12.0pt}
  \caption{Left: Percentage of flips in gender (CLIP) from editing synthetic (SD) Male and Female images using high-paid prompts in diffusion-based image editing (across different levels of edit strength). Middle: Skin Color Changes ($\uparrow$ change towards lighter skin color using an established methodology described in \Sref{sec:methodlogy}) from editing synthetic (SD) images of White and Black individuals using high-paid prompts in diffusion-based image editing (across different levels of edit strength). Right: Percentage of diffusion-based classifier choices towards male-dominated professions in binary classification tasks between a male- and female-dominated profession pair (at different numbers of noise samples in the estimation of the classification objective) for synthetic (SD) images.}
  \vspace{-9.0pt}
  \label{fig:BiasDownstreamGraph-SYN}  
\end{figure}

\end{document}